\pdfoutput=1
\documentclass[fleqn,usenatbib]{mnras}

\usepackage[T1]{fontenc}
\usepackage{ae,aecompl}

\usepackage{graphicx}	% Including figure files
\usepackage{amsmath}	% Advanced maths commands
\usepackage{amssymb}	% Extra maths symbols
\usepackage{caption}
\usepackage{longtable}
\usepackage{newtxtext,newtxmath}
\usepackage{booktabs}
\usepackage{morefloats}
\usepackage{footnote}
\usepackage{multirow}
\usepackage{color}
\definecolor{referee}{gray}{0}
\usepackage{pdflscape}

\title[HerBS Sample - SCUBA-2 Observations]{The \textit{Herschel} Bright Sources (HerBS): \\Sample definition and SCUBA-2 observations.}

\author[Tom Bakx]{Tom J. L. C. Bakx$^{1}$\thanks{E-mail: bakxtj@cardiff.ac.uk (Cardiff)},
S. A. Eales$^{1}$,
M. Negrello$^{1}$, 
M. W. L. Smith$^{1}$,
E. Valiante$^{1}$,\newauthor 
W. S. Holland$^{2}$,
M. Baes$^{3}$,
N. Bourne$^{4}$,
D. L. Clements$^{5}$,
H. Dannerbauer$^{6,20,21}$,\newauthor
G. De Zotti$^{7}$,
L. Dunne$^{1}$,
S. Dye$^{8}$,
C. Furlanetto$^{8,9}$,
R. J. Ivison$^{4,10}$,
S. Maddox$^{1}$,\newauthor
L. Marchetti$^{11}$,
M. J. Micha{\l}owski$^{4,19}$,
A. Omont$^{12,13}$,
I. Oteo$^{4,10}$,
J. L. Wardlow$^{14}$,\newauthor
P. van der Werf$^{15}$ and
C. Yang$^{12,13,16,17,18}$.
\\
$^{1}$School of Physics \& Astronomy,  Cardiff University, The Parade, Cardiff, CF24 3AA UK\\
$^{2}$UK Astronomy Technology Centre, Royal Observatory, Blackford Hill, Edinburgh EH9 3HJ, UK\\
$^{3}$Sterrenkundig Observatorium, Universiteit Gent, Krijgslaan 281 S9, B-9000 Gent, Belgium\\
$^{4}$Institute for Astronomy, University of Edinburgh, Royal Observatory, Edinburgh EH9 3HJ, UK\\
$^{5}$Astrophysics Group, Imperial College, Blackett Laboratory, Prince Consort Road, London SW7 2AZ, UK\\
$^{6}$Universit{\"a}t Wien, Institut f{\"u}r Astrophysik, T{\"u}rkenschanzstrabe 17, 1180 Wien, Austria\\
$^{7}$INAF-Osservatorio Astronomico di Padova, Vicolo dell'Osservatorio 5, I-35122 Padova, Italy\\
$^{8}$School of Physics and Astronomy, University of Nottingham, University Park, Nottingham NG7 2RD, UK\\
$^{9}$CAPES Foundation, Ministry of Education of Brazil, Bras\'ilia/DF, 70040-020, Brazil\\
$^{10}$European Southern Observatory, Karl-Schwarzschild-Strasse 2, 85748 Garching bei M{\"u}nchen, Germany\\
$^{11}$Department of Physical Science, The Open University, Milton Keynes MK7 6AA, UK\\
$^{12}$UPMC Univ. Paris 06, UMR7095, Institut d'Astrophysique de Paris, 75014 Paris, France\\
$^{13}$CNRS, UMR7095, Institut d'Astrophysique de Paris, 75014 Paris, France\\
$^{14}$Centre for Extragalactic Astronomy, Department of Physics, Durham University, South Road, Durham, DH1 3LE, UK\\
$^{15}$Leiden Observatory, Leiden University, P.O. Box 9513, NL - 2300 Leiden, The Netherlands\\
$^{16}$Purple Mountain Observatory/Key Lab of Radio Astronomy, Chinese Academy of Sciences, Nanjing 210008, PR China\\
$^{17}$Institut d'Astrophysique Spatiale, CNRS, Univ. Paris-Sud, Universit{\'e} Paris-Saclay, Bât. 121, 91405 Orsay Cedex, France\\
$^{18}$Graduate University of the Chinese Academy of Sciences, 19A Yuquan Road, Shijingshan District, 10049, Beijing, PR China \\
$^{19}$Astronomical Observatory Institute, Faculty of Physics, Adam Mickiewicz University, ul.~S{\l}oneczna 36, 60-286 Pozna{\'n}, Poland\\
$^{20}$Instituto de Astrofísica de Canarias (IAC), E-38205 La Laguna, Tenerife, Spain\\
$^{21}$Universidad de La Laguna, Dpto. Astrof{\'i}sica, E-38206 La Laguna, Tenerife, Spain
}

\date{Accepted XXX. Received YYY; in original form ZZZ}

\pubyear{2017}

\begin{document}
\label{firstpage}
\pagerange{\pageref{firstpage}--\pageref{lastpage}}
\maketitle

\begin{abstract}
\noindent
We present the \textit{Herschel} Bright Sources (HerBS) sample, a sample of bright, high-redshift \textit{Herschel} sources detected in the 616.4 square degree H-ATLAS survey. The HerBS sample contains 209 galaxies, selected with a 500 $\mu$m flux density greater than 80 mJy and an estimated redshift greater than 2. The sample consists of a combination of HyLIRGs and lensed ULIRGs during the epoch of peak cosmic star formation.
In this paper, we present SCUBA-2 observations at 850 $\mu$m of 189 galaxies of the HerBS sample, 152 of these sources were detected.
We fit a spectral template to the \textit{Herschel}-SPIRE and 850 $\mu$m SCUBA-2 flux densities of 22 sources with spectroscopically determined redshifts, using a two-component modified blackbody spectrum as a template. We find a cold- and hot-dust temperature of  $21.29_{-1.66}^{+1.35}$ K and $45.80_{-3.48}^{+2.88}$ K, a cold-to-hot dust mass ratio of $26.62_{-6.74}^{+5.61}$ and a $\beta$ of $1.83_{-0.28}^{+0.14}$. The poor quality of the fit suggests that the sample of galaxies is too diverse to be explained by our simple model.
Comparison of our sample to a galaxy evolution model indicates that the fraction of lenses is high. Out of the 152 SCUBA-2 detected galaxies, the model predicts 128.4 $\pm$ 2.1 of those galaxies to be lensed (84.5\%). The SPIRE 500 $\mu$m flux suggests that out of all 209 HerBS sources, we expect 158.1 $\pm$ 1.7 lensed sources, giving a total lensing fraction of 76 per cent. 
\end{abstract}


%%%%%%%%%%%%%%%%%%%%%%%%%%%%%%%%%%%%%%%%%%%%%%%%%%

%%%%%%%%%%%%%%%%% BODY OF PAPER %%%%%%%%%%%%%%%%%%

%	ARTICLE CONTENTS
%----------------------------------------------------------------------------------------
\section{Introduction} % The \section*{} command stops section numbering

\addcontentsline{toc}{section}{Introduction} % Adds this section to the table of contents

The \textit{Herschel Space Observatory} \citep{Pilbratt2010} has increased the number of known sub-millimetre galaxies (SMGs) from hundreds to hunderds of thousands. The H-ATLAS survey (\textit{Herschel} Astrophysical Terahertz Large Area Survey - \citealt{Eales2010}; \citealt{Valiante2016}) is one of the largest legacies of \textit{Herschel}. This survey observed a total of 616.4 square degrees over five fields in five wavebands. The large-area surveys done with \textit{Herschel} allow us to select sources that are among the brightest in the sky, of which a large percentage are lensed ULIRGs (Ultra-Luminous Infrared Galaxies, $10^{12}$ L$_{\odot}$ < L$_{\rm{FIR}}$ < $10^{13}$ L$_{\odot}$) and HyLIRGs (Hyper-Luminous Infrared Galaxy, L$_{\rm{FIR}}$ > $10^{13}$ L$_{\odot}$) at high redshift.

A similar selection for bright sources was already exploited in the 14.4 sqr. deg. Science Demonstration Phase (SDP) of H-ATLAS by \cite{Negrello2010}, who used a simple flux cut-off to select lensed sources. They were able to remove all contaminants from their selection, local galaxies and blazars, and identified five lensed galaxies.
\cite{Wardlow2013} followed a similar approach on the 94 sqr. deg. HerMES (\textit{Herschel} Multi-tiered Extragalactic Survey) maps, and selected 13 sources with S$_{\rm{500\mu m}}$ > 100mJy. Nine of these sources had follow-up data, done with the Sub-Millimetre Array (SMA), the Hubble Space Telescope (HST), Jansky Very Large Array (JVLA), Keck, and Spitzer. \cite{Wardlow2013} combined these data for six sources and confirmed their lensing nature, while three other sources had their lensing nature already confirmed by \cite{Borys2006}, \cite{Conley2011}, and \cite{Ikarashi2011}.
Recently, \cite{Negrello2016} and \cite{Nayyeri2016} used the same S$_{\rm{500\mu m}}$ > 100mJy flux density cut-off on the full H-ATLAS (616.4 sqr. deg.) and HeLMS \textcolor{referee}{(HerMES Large Mode Survey};  372 sqr. deg.) maps, and created samples containing 77 and 80 sources, respectively. Spectroscopic and optical follow-up observations were able, so far, to confirm that 20 sources are indeed lensed, one is a proto-cluster \citep{Ivison2013}, while the remaining sources in \cite{Negrello2016} await more observations to be carried out to confirm their nature.

Large samples of lensed sources are interesting, both because of the lensed source and the intervening lensing galaxy \citep{Treu2010}.
The lensed source has an amplified flux density and increased angular size. 
The amplification in flux density allows us to study sources that would otherwise be too faint to detect.
The increase in angular size allows us to study the internal properties of high redshift sources with high resolution sub-mm/mm and radio observatories, such as ALMA (Atacama Large Millimeter Array) and the VLA (Very Large Array).
As most intervening, lensing sources are passively evolving ellipticals, they are sub-mm dim and their contribution to the total measured flux density is minimal.
This allowed \cite{ALMA2015}, \cite{Dye2015}, \cite{Hatsukade2015}, \cite{Rybak2015}, \cite{Swinbank2015} and \cite{Tamura2015} to study SDP.81 down to sub-kiloparsec scales, using the increase in angular size in order to resolve the morphological and dynamical properties of a galaxy at a redshift of 3.

Sub-mm detected lensed sources, similar to SDP.81, are forming stars at rates of hundreds to several thousands of solar masses per year, and large samples of them can allow statistically significant studies into these extremely star-forming sources. 
This is important, because the comoving density of ULIRGs at z = 2 to 4 is about a thousand times greater than in the local universe, and these dusty star-forming galaxies are estimated to contribute about 10\% of the total star formation in this redshift range \citep{Hughes1998, Blain1999, Smail2002,Wardlow2011,Casey2014}. This means that SMGs contribute significantly to the peak in cosmic star formation, which occurred around z $\sim$ 2.3 \citep{Chapman2005}.

While the star-formation rate of the universe has been measured up to redshift z $\sim$ 8 in rest-frame UV surveys, these studies only measure the unobscured star-formation rates \citep{Madau2014}. The star formation processes in these dusty star-forming galaxies (DSFGs) tend to be obscured by the dust, and are missed by current optical investigations of the cosmic star-formation rate.
An added benefit of using sub-mm observations to measure the obscured star-formation rate is that sub-mm flux density falls only slowly with redshift, because of the negative K-correction:
sub-mm observations observe the Rayleigh-Jeans part of the modified blackbody spectrum, which causes the flux density to increase as the galaxy's redshift increases. This increase is able to compensate for the cosmological dimming due to the increase of luminosity distance, e.g. a redshift 1 or 4 galaxy has a similar flux density in sub-mm wavelengths \citep{Blain1993,Blain2002,Bethermin2015}.

The foreground galaxy's total mass (dark and baryonic) distribution determines the lensed morphology of the sub-mm detected system \cite{Vegetti2012,Hezaveh2016a,Hezaveh2016b}.
Therefore, high-resolution imaging of the lensed morphology allows the detection of low-mass substructures in lensing galaxies.
These substructures can then be used to test the formation of structure in large-scale cosmological simulations, such as the Millennium \citep{Springel2005} and the recent Eagle simulation \citep{Schaye2015}.

The statistics of galaxy-galaxy lensing systems furthermore allows for a measurement of global cosmological parameters.
For example, the lensing statistics of 28 lensed quasars in the Sloan Digital Sky Survey (SDSS) Quasar Lens Search (SQLS) gave an estimate of $\Omega_{\Lambda}$ = 0.74 $\pm$ 0.17, assuming a spatially flat universe \citep{Oguri2012b}.
Selecting lensed sources from bright sub-mm samples is simple and unbiased method because it is based on the source, as the lens is usually faint in the sub-mm.
\cite{Eales2015} showed that observations of a sample of 100 lensed \textit{Herschel} sources would be enough to estimate $\Omega_{\Lambda}$ with a precision of 5 per cent and observations of 1000 lenses would be enough to estimate $\Omega_{\Lambda}$ with a precision similar to that obtained from the \textit{Planck} observations of the cosmic microwave background.

A high flux density cut-off (S$_{\rm{500\mu m}}$ > 100 mJy) eliminates a large amount of possible lenses in order to achieve a low contamination rate from unlensed sources \citep{GN2012}.
Lowering the cut-off flux density to 80 mJy was already tested in \cite{Wardlow2013}. Out of the four galaxies with lensing verification, only one was confirmed to be a lens.
In this paper, we will reinvestigate the question of using a lower cut-off flux, by selecting galaxies from the 616.4 sqr. deg. H-ATLAS survey. In order to decrease the contamination rate, we impose a photometric cut-off redshift z$_{\rm{phot}}$ $>$ 2 based on the \textit{Herschel}-SPIRE fluxes. The probability of lensing below this redshift falls off sharply, because of the smaller volume available between us and the source \citep{Strandet2016}. We will calculate the expected amount of lensed galaxies in our sample, by comparing the fluxes of our sources to a cosmological evolution model that takes lensing into account.

Our sample selection is based on \textit{Herschel} fluxes, and a known problem of sources selected at 500 $\mu$m with Herschel is the large solid angle of the beam \citep{Scudder2016}. This could lead to several sources blending into a single source, and result in a flux that is too large. This is why we observed the majority of our sources at 850 $\mu$m with the SCUBA-2 instrument on the James Clerk Maxwell Telescope (JCMT), whose beam has a six times smaller solid angle on the sky. The extra data point should also improve the photometric redshift estimates of our sources. 

In Section \ref{sec:measurements}, we discuss the selection of the \textit{Herschel} Bright Sources (HerBS) sample, as well as the observations with SCUBA-2. We describe the results of the JCMT observations in Section \ref{sec:results}, where we also remove several blazar contaminants from the sample. We re-derive a spectral template for our sources with spectroscopically determined redshifts in Section \ref{sec:data}. We discuss the effects of source confusion, the properties of the template, the redshift distribution of our sample, and estimates of the lensing fraction in Section \ref{sec:discussion}.

Throughout this paper we assume the $\Lambda$-CDM model, and the best-fit parameters found by the \cite{Planck2015}: $H_{\rm{0}}$ 67.7 km s$^{-1}$ Mpc$^{-1}$ and $\Omega_{\rm{M}}$ = 0.307.

%------------------------------------------------

% ++++++ METHODS ++++++
\section{Sample and measurements}
\label{sec:measurements}

\begin{table*}
	\centering
	\caption{The H-ATLAS fields}
	\label{tab:HATLAS}
	\begin{tabular}{lccccccc}
	\hline
		Field 	& \multicolumn{2}{c}{Centre} 	& \multicolumn{2}{c}{Approximate dimensions} & Final surface area & Sources & Surface density\\
		 		& RA [hms]	&	DEC	[dms]		&	RA [deg]	& DEC [deg]	& [sqr. deg.] & & [1/sqr. deg]\\
		\hline
		\textbf{NGP} 	& 13:18:00  & 29:00:00 	& 15 	& 10 	& \textbf{170.1} & \textbf{49} & \textbf{0.288}\vspace{0.1cm}\\
		\textbf{GAMA Total}	&	-	&	-		&	-	&	-	& \textbf{161.6} & \textbf{72} & \textbf{0.446}\vspace{0.0cm}\\
		GAMA 9 	& 09:00:00	& 00:00:00	& 12	& 3		& 53.43 & 23 & 0.430\\
		GAMA 12 	& 12:00:00	& 00:00:00	& 12	& 3		& 53.56 & 26 & 0.485\\
		GAMA 15 	& 14:30:00	& 00:00:00	& 12	& 3		& 54.56 & 23 & 0.422\vspace{0.1cm}\\
		\textbf{SGP} 	& 23:24:46	& -33:00:00	& 42	& 6		& \textbf{284.8} & \textbf{88} & \textbf{0.309}\\
		\hline
		\textbf{Total fields}	&		-	&	-		&	-	&	-	& \textbf{616.4} & \textbf{209} & \textbf{0.339}
	\end{tabular}
	\\ \flushleft \vspace{0.2cm} \textbf{Notes:} Reading from the left, the columns are: Column 1 - name of field; Column 2 and 3 - The location of the centre of the field; Column 4 and 5 - The approximate dimensions of the field; Column 6 - The surface area from the final maps \citep{Valiante2016}; Column 7 - The number of final HerBS sources in each field; Column 8 - The surface density of HerBS sources per field.
\end{table*}

\begin{figure*}
  \centering
  \includegraphics[width=\linewidth]{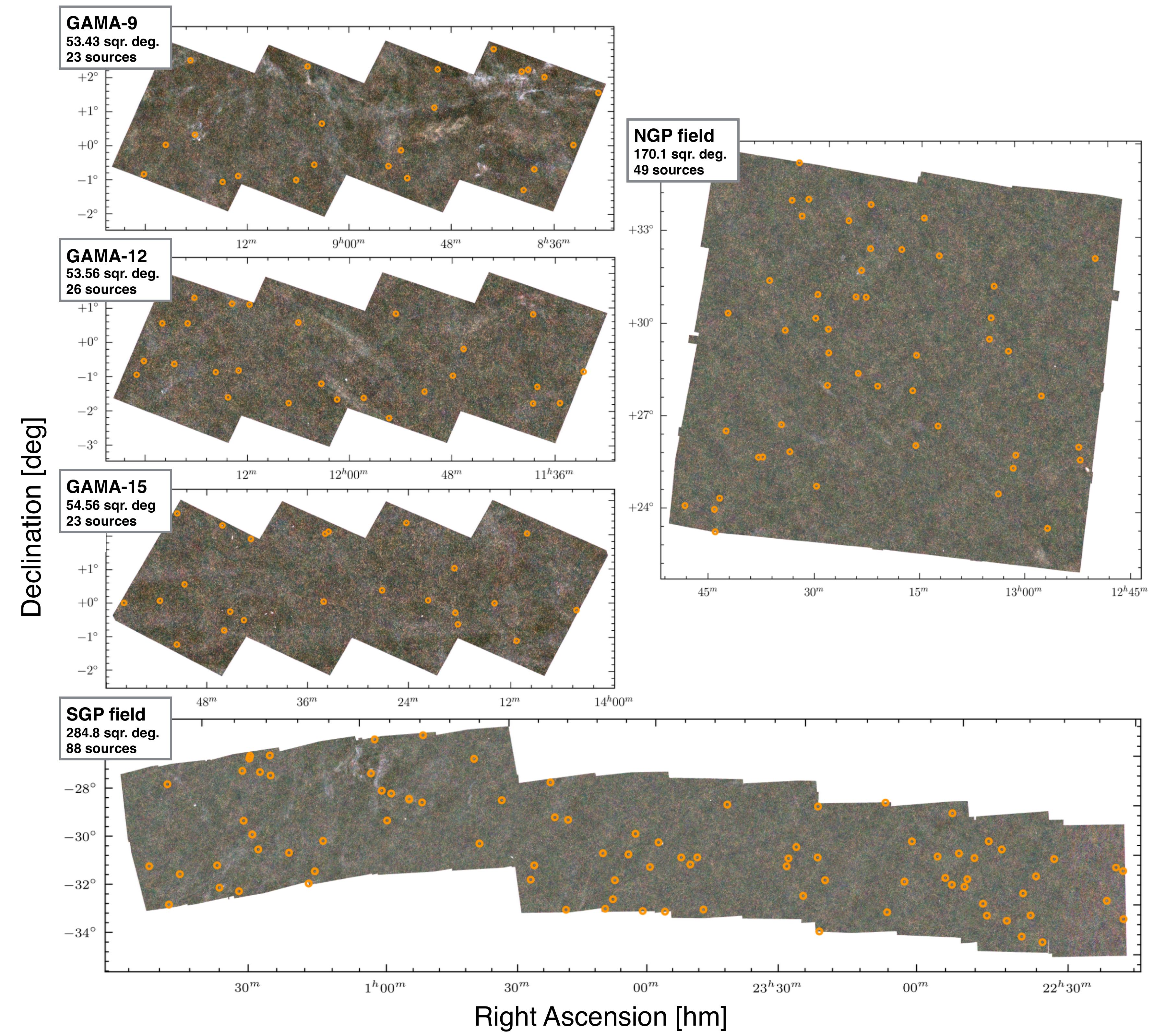}
  \caption{\textit{Herschel}/SPIRE color maps of the H-ATLAS fields. The orange circles mark the positions of the 209 HerBS sources. This figure is similar to Figure 2 in \citep{Negrello2016}, and shows how the sources are distributed over the sky.}
  \label{fig:sourcesTogether}
\end{figure*}

\subsection{The selection of the HerBS sample}
The sample was selected from the brightest, high-redshift sources in the H-ATLAS survey. The H-ATLAS survey used the PACS \citep{Poglitsch2010} and SPIRE \citep{Griffin2010} instruments on the \textit{Herschel Space Observatory} to observe the North and South Galactic Pole Fields and three equatorial fields to a 1$\sigma$ sensitivity of 5.2 mJy at 250 $\mu$m to 6.8 mJy at 500 $\mu$m, although the noise varies per source \citep{Valiante2016}. The three equatorial fields overlap with the Galaxy And Mass Assembly (GAMA) fields 9, 12 and 15 hours, and from here on we adopt this naming convention for the equatorial fields \citep{Driver2011,Liske2015}. 
The fields are defined in Table \ref{tab:HATLAS}. In total the H-ATLAS survey detected approximately half a million sources.

We initially selected the HerBS sample from the H-ATLAS point-source catalogues \citep{Valiante2016}, \textcolor{referee}{who extracted the flux densities at the 250 $\mu$m position, and used this position for flux extraction at 350 and 500 $\mu$m.} The flux densities in the catalogues are not de-boosted, however the flux boosting is negligible compared to the flux uncertainty; around 1 per cent at 80 mJy, and diminishing for increasing flux density, as can be seen in Table 6 of \cite{Valiante2016}.
We estimated the redshift of each source by fitting a source template to the 250, 350 and 500 $\mu$m flux densities \citep{Pearson2013}. 
We selected the sources at an estimated redshift, $z_{\rm{phot}}$, greater than 2 and a 500 $\mu$m flux density, S$_{\rm{500\mu m}}$, greater than 80 mJy. 
The source template was a two-temperature modified blackbody from \cite{Pearson2013} (see eq. \ref{eq:BB} and Table \ref{tab:subfitting} in our Section \ref{sec:data}).
This modified blackbody was derived from the \textit{Herschel} PACS and SPIRE flux densities of 40 sources with spectroscopically determined redshifts, with 25 sources at low redshifts (z $<$ 1), and only 12 sources at high redshifts (z $>$ 2).
Our initial sample consisted of the 223 sources. 

Where possible we removed sources that are coincident with a large nearby galaxy or  a blazar \citep{Negrello2010, Lopez-Caniego2013}. However, the preselection of blazars was not complete, and it only became clear after we had carried out the SCUBA-2 observations that we had actually observed several blazars (see Section \ref{sec:results}).  
The final HerBS sample consists of 209 sub-millimetre galaxies after removing all nearby galaxies and blazars, and is listed in Table \ref{tab:AppendixA}. 
We plot the positions of the final 209 HerBS sources in the various fields in Figure \ref{fig:sourcesTogether}.

Several of the HerBS sources have been investigated individually. \cite{Fu2012} showed that  HATLAS J114637.9-001132 (HerBS-2) is a strongly lensed sub-mm galaxy, with a magnification between 7 to 17. \cite{Cox2011} and \cite{Bussmann2012} found that HATLAS J142413.9+022303 (HerBS-13) is a lensed sub-mm galaxy, with a magnification of 4. At a redshift of 4.24, the source has one of the highest redshifts in our sample. HATLAS J090311.6+003907 (HerBS-19) is also known as SDP.81, and has recently been observed by \cite{ALMA2015}. \cite{Negrello2010} showed SDP.81 is lensed using 880 $\mu$m Sub-Millimetre Array observations. \cite{Dye2015} and \cite{Tamura2015} reconstructed the galaxy from the ALMA observation, by modelling the distorting effect of the lens. They found a magnification of $\sim$ 11. This reconstructed image features details on the scale of hundreds of parsecs, and the image shows resolved individual giant molecular clouds in a z = 3.04 galaxy. SDP.81 appears, through reconstructed HST and ALMA imaging, to be two interacting objects, where the dust disk is in a state of collapse.

However, not all our sources are lensed. \cite{Ivison2013} studied HATLAS J084933.4+021442 (HerBS-8), and found it was not a strongly lensed galaxy. Instead, it consists of multiple large galaxies in the process of merging, which has probably triggered starbursts in the individual galaxies, explaining the brightness in sub-mm wavelengths.

Our HerBS sample overlaps partially with the sample from \cite{Negrello2016}, as 53 out of the 80 sources in their sample are also found in the HerBS sample. Their sample was designed specifically to find lensed systems, by imposing a flux-density cut-off at 100 mJy at 500 $\mu$m and did not have a lower redshift limit.

\subsection{Observations with SCUBA-2}
We observed 203 sources with the SCUBA-2 array on the JCMT. The instrument consists of 10,000 Transition Edge Sensor (TES) bolometers, distributed over 4 arrays that observe at 450 $\mu$m and 4 arrays that observe at 850 $\mu$m \citep{Holland2013}. Both wavelengths are observed simultaneously, with the use of a dichroic mirror. The voltage across each array is optimised to ensure as many functional bolometers as possible. The optimised voltage places the majority of the bolometers within their sensitive resistance transition, whereupon any temperature fluctuation causes a current change. The resulting magnetic field variations are read out with separate Superconducting Quantum Interference Devices (SQUIDs) located under each bolometer.

The instrument scans the sky in a DAISY pattern, circling around the source following a continuous petal-like track, providing a central 3 arc-minute region of uniform exposure time, and keeping one part of the array on-source at all times \citep{Chapin2013}.

The observations conditions were in the grade-3 weather band [0.08 $<$ $\tau_{\rm{1.3 mm}}$ $<$ 0.12], which is only suitable for 850 $\mu$m observations. The data were flux-calibrated against Uranus, Mars, CRL 618 and CRL 2688 (the Westbrook and Egg Nebulae). The calibrators were observed between 2 and 4 times per observing run, and the flux calibration factors (FCFs) were estimated linearly for observations in between calibrators, and the closest calibrator was used otherwise \citep{Dempsey2013}.

Our observations consisted of ten-minute exposures for each source. The bolometers are sampled at roughly 200 Hz, and the data is stored in 30-second time slices for each of the arrays, where the first and last time slice of each exposure are flat-fields. Flat-fields probe the responsivity of individual bolometers, and are derived from the bolometer's response to the resistance heaters, which are located next to each bolometer. 

\subsection{Data reduction}
\begin{figure}
  \centering
  \includegraphics[height=0.87\textheight]{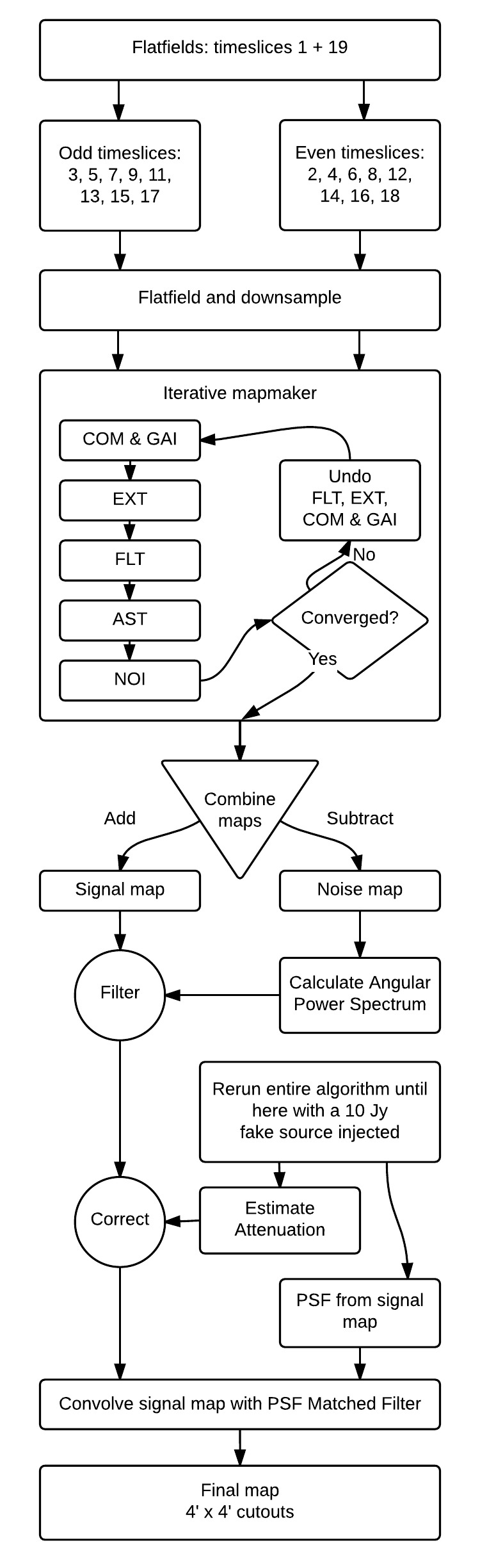}
  \caption{This flowchart shows the data reduction steps schematically, starting from the raw data files at the top, working to the reduced cutouts at the bottom. The intricacies are detailed in the data reduction section. For each observation, two sets of timeslices are cleaned and processed through the iterative mapmaker, and these resulting maps are subtracted to provide a jackknife estimate of the noise. A fake source is injected to estimate peak attenuation due to the filtering process, and allows us to create a PSF for the final matched filter step.}
  \label{fig:datareduction}
\end{figure}

The entire data reduction method is shown schematically in Figure \ref{fig:datareduction}, and is described below. The data reduction was done with the ORAC$\_$DR pipeline, which uses the KAPPA and SMURF packages from STARLINK, and the PICARD procedures \citep{Council2014}.

The basic data consists of the time-dependent signals from each bolometer and information about the specific scanning pattern of the arrays on the sky. The first step of the data reduction method flat-fields and downsamples the data, to correct for individual bolometer performance and to reduce the file size by matching the sampling speed to the spatial scale of the maps. The second step removes the noise components in the signal iteratively, starting with the largest noise component \citep{Chapin2013}. Our final reduced map is achieved with additional data reduction steps: jackknife, fake point-source injection and matched filtering. The final result is a 4 by 4 arcminute image with one arcsecond resolution.

\subsubsection*{The iterative data reduction step ($makemap$)}
Sky emission is the dominant noise component, and it is shared by all bolometers. This common-mode signal (COM) is calculated by averaging the signals of all bolometers into one signal per subarray. The common-mode signal is then subtracted from the signal for each bolometer, taking care to adjust for individual bolometer amplification differences (GAI). Bolometers that have a signal that is inconsistent with the common-mode signal are rejected at this stage. 

The signal is then corrected for the atmospheric extinction (EXT), a function of precipitable water vapour and telescope pitch, after which a high-pass Fourier filter (FLT) removes low-frequency, 1/f noise. The frequency cut-off is 0.8 Hz, which corresponds to a spatial scale of 200 arc-seconds.

The next step removes the astronomical signal (AST) from the total signal, in order to estimate convergence of our iterative data reduction step. The signals of all bolometers are projected onto the sky, creating an astronomical map of our observation. Many data points contribute to the estimate of the astronomical signal in each spatial pixel, which greatly reduces the noise compared to the time-series data. The map still contains noise, but the assumption made in this step of the iterative data-reduction procedure is that everything in this map is real. The astronomical, space-domain map is then used to create a time-domain signal for each bolometer, by simulating an observation of our astronomical map. This is then removed from the signal for each bolometer.

The time-domain signal for each bolometer should now consist only of noise. This noise is calculated and compared to the convergence criterion (NOI), which is a minimum number of loops (four in this case) and a threshold noise level. If convergence is not reached in the NOI step, all the data-processing steps (FLT, EXT, GAI, COM) are undone, except for the removal of the astronomical signal. This adds back the common-mode noise and the noise removed in the Fourier-filtering step. All the steps (see upper half of Figure \ref{fig:datareduction}) are then repeated until the convergence criterion is met. After each cycle the new estimate of the astronomical signal is added to the previous estimate. The final image is obtained when the convergence criterion is met.

\subsubsection*{Extra data reduction steps}
Apart from this standard data-reduction procedure, shown in the top half of Figure \ref{fig:datareduction}, we added the following additional steps.

For each source, we split the time-slices into two sets. Each set consists of the flat-fields (first and last time slice) and either the odd or even half of the time slices. We ran the iterative mapmaker over each set, separately, which allows us to execute a jackknife step (ORAC$\_$DR procedure: SCUBA2$\_$JACKKNIFE). 

We used the iterative data reduction step to create a separate map for each half of the data. We subtracted one map from the other to create a noise-map, from which we calculate the angular power spectrum of the noise. We used this angular power spectrum to construct a map-specific Fourier filter. A combined signal map is calculated by adding the two signal maps, and we then applied this Fourier-filter to the signal map.

The high-pass filtering step attenuates the signal, and to account for this, we reran the entire data reduction algorithm with an injected fake source. This fake 10 Jy point-source (FWHM of 13 arc seconds - the main beam size of 850 $\mu$m observations with JCMT \citep{Dempsey2013}) was injected into both the odd and even timeslices, offset at 30 arc seconds from the centre. This extremely bright, fake source allowed us to calculate an effective point spread function (PSF) and also provided an estimate of the signal attenuation due to the high-pass filtering, which usually was around 15 to 20\%.

Finally, we applied a matched filter to the signal map, in which we convolved our signal map with the PSF found by injecting a fake source. This provided the final, reduced observation map. We cropped the observation to a 4 by 4 arcminute image, and measured the fluxes by measuring the highest flux density pixel in the central 50 by 50 arcsecond region around the SPIRE-estimated position. We determine a SCUBA-2 detection by a combination of proximity to the \textit{Herschel}-SPIRE 250$\mu$m position and the signal to noise, as shown in Section \ref{sec:results}.

% ++++++ Results ++++++
\section{Results}
\label{sec:results}

We observed 203 of our preselected H-ATLAS sources with the SCUBA-2 instrument. In the following analysis, we find that fourteen detected sources turn out to be blazars, which leaves our entire HerBS galaxy sample containing 209 sources. 152 of these sources are detected, 27 sources are not detected due to a signal-to-noise cut, and ten sources do have a 3-$\sigma$ detection, but not within the 10 arcsecond circle around the SPIRE position. These results are summarised in Table \ref{tab:Observations}. 

\begin{figure}
	\includegraphics[width=\linewidth]{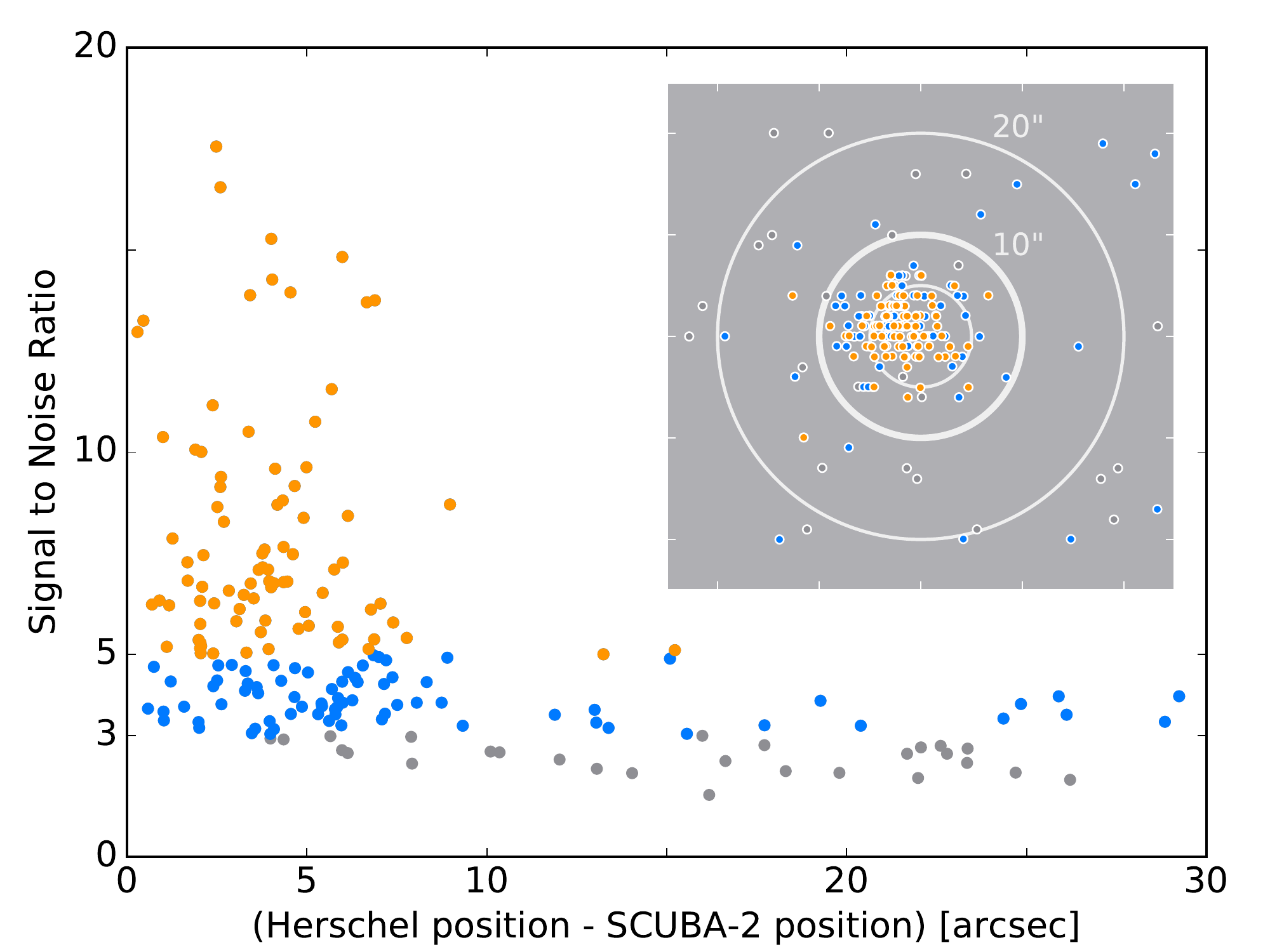}
	\caption{The majority of high signal-to-noise SCUBA-2 fluxes lie in a 10 arcsecond circle around the SPIRE position. We choose a cut-off signal-to-noise ratio of 3-$\sigma$, and a maximum radius of 10 arcseconds. The fifteen sources with a signal-to-noise ratio between 3 and 5 suggest that the HerBS sources might have two false detections. The overlay graph shows the position of the SCUBA-2 observation, where each point was centered on the SPIRE position.}
	\label{fig:positionDistribution}
\end{figure}

Figure \ref{fig:positionDistribution} shows the distribution of the maximum signal to noise in a 50 by 50 arcsecond box centered on the SPIRE position, as a function of the position offset. 

We decide to define a detected source by a signal-to-noise greater than 3 and a positional offset smaller than 10 arcseconds. Initially, we find 159 sources that satisfy this criterion, 27 sources that are not detected by the signal-to-noise cut, and 17 sources whose positional offset was too large.

For each of the seventeen sources that do not have their maximum flux within the 10 arcsecond circle around the SPIRE position, that do have a signal-to-noise greater than 3, we decreased the size of the searching box to find the peak in flux. Of these seventeen sources, seven sources have fluxes within 10 arc seconds from the SPIRE position with a signal-to-noise greater than 3, as show in boldface in Table \ref{tab:new850s}. These seven sources are added to the detected sources.

Of the sources with signal to noise ratios between three and five, fifteen are originally situated outside of the 10 arcsecond circle. These sources are distributed over 89 per cent of the map (the area outside the 10 arcsecond circle). An even distribution of such false detections would result in two ($\sim1.7$) false detections inside the HerBS catalogue. The overlay graph inside Figure \ref{fig:positionDistribution} shows a strong correlation for most points around the centre, however all other non-detections appear uniformly scattered, making an even distribution likely.

\begin{figure}
	\includegraphics[width=\linewidth]{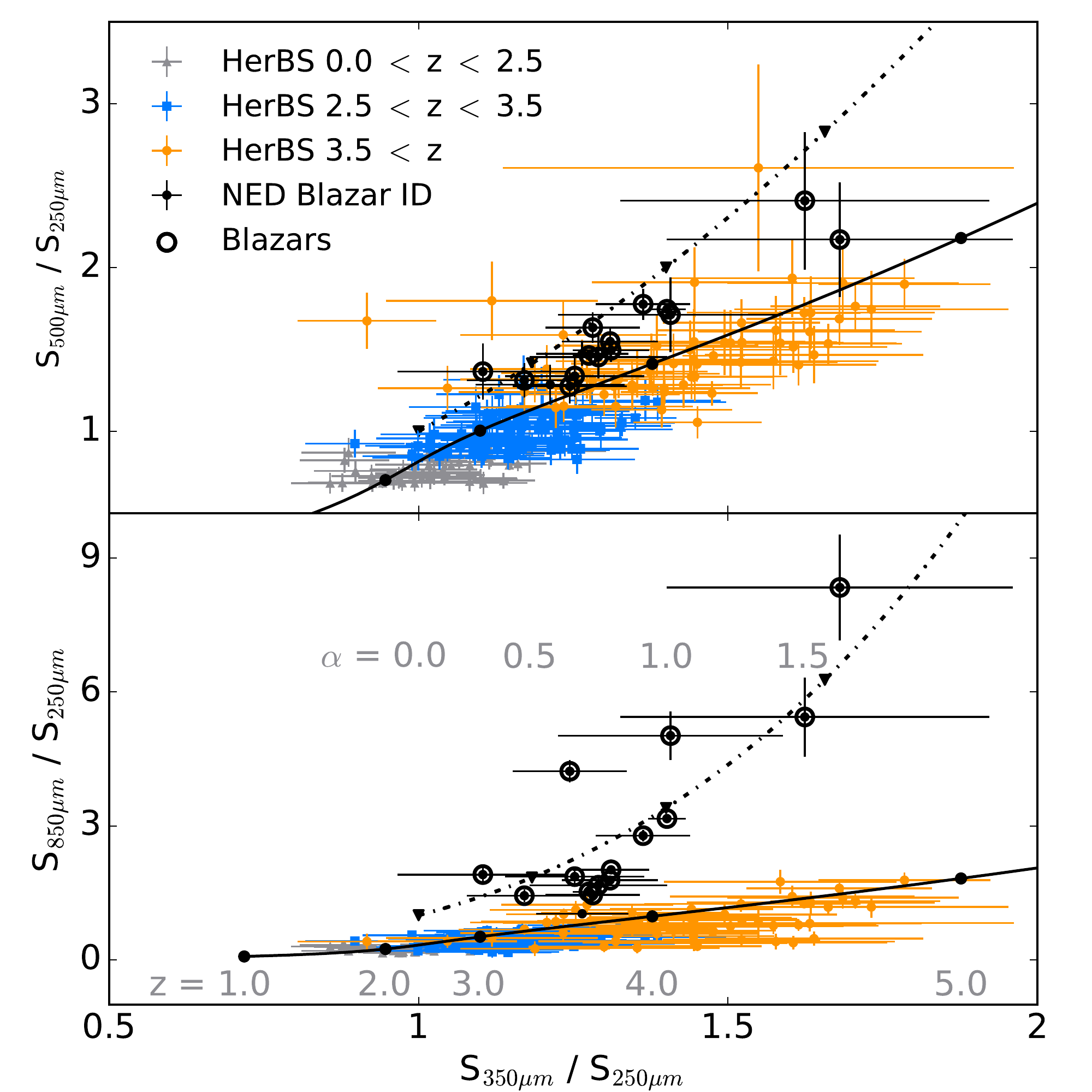}
	\caption{The top panel shows the flux ratios based on just \textit{Herschel} fluxes. We plot S$_{\rm{500\mu m}}$/S$_{\rm{250\mu m}}$ versus S$_{\rm{350\mu m}}$/S$_{\rm{250\mu m}}$. Sources close to a known blazar in NED (\textit{black circles}) lie in the same region as the high-redshift HerBS sources (\textit{gray triangles, blue squares and red circles}). The bottom panel shows the flux ratios when we include the SCUBA-2 observations. We plot S$_{\rm{850\mu m}}$/S$_{\rm{250\mu m}}$ against S$_{\rm{350\mu m}}$/S$_{\rm{250\mu m}}$. Most sources close to a known blazar occupy a different region of the graph, and can be easily identified and removed (\textit{black circles}). The difference between the graphs indicates the necessity of the 850 $\mu$m observations for removing blazar contaminants from the sample. We also plot the track for the template we derive in Section \ref{sec:template} through the diagram as the redshift changes (\textit{black line and circles}). Similarly, we show the expected blazar track, \textcolor{referee}{for alpha-values ranging from 0 to 1.5 (\textit{black dash-dot line and triangles})}.}
	\label{fig:500vs850}
\end{figure}

We know from \cite{Negrello2007} that there is a risk that several of these sources are blazar contaminations. In order to find these contaminants, we plot their flux ratios in Figure \ref{fig:500vs850}.

The top panel shows the flux ratios based on just \textit{Herschel} fluxes. We plot S$_{\rm{500\mu m}}$/S$_{\rm{250\mu m}}$ versus S$_{\rm{350\mu m}}$/S$_{\rm{250\mu m}}$. The sources that lie very close to a known blazar \textcolor{referee}{(within 10 arc seconds)} in the NASA Extragalactic Database (NED) (\textit{black circles}) lie in the same region as the high-redshift HerBS sources (\textit{gray triangles, blue squares and red circles}). We also plot the track for the template we derive in Section \ref{sec:template} through the diagram as the redshift changes (\textit{black line and circles}). Similarly, we show the expected blazar track (assuming synchrotron radiation), for various possible alpha-values (\textit{black dash-dot line and triangles}). Note that both these tracks do not differ significantly from each other. The bottom panel shows the flux ratios of the 203 sources with SCUBA-2 observations. We plot S$_{\rm{850\mu m}}$/S$_{\rm{250\mu m}}$ against S$_{\rm{350\mu m}}$/S$_{\rm{250\mu m}}$. 
Most of the galaxies close to a known blazar occupy a different region of the graph, and can be easily identified and removed from the sample. 

\begin{figure*}
	\includegraphics[width=\linewidth]{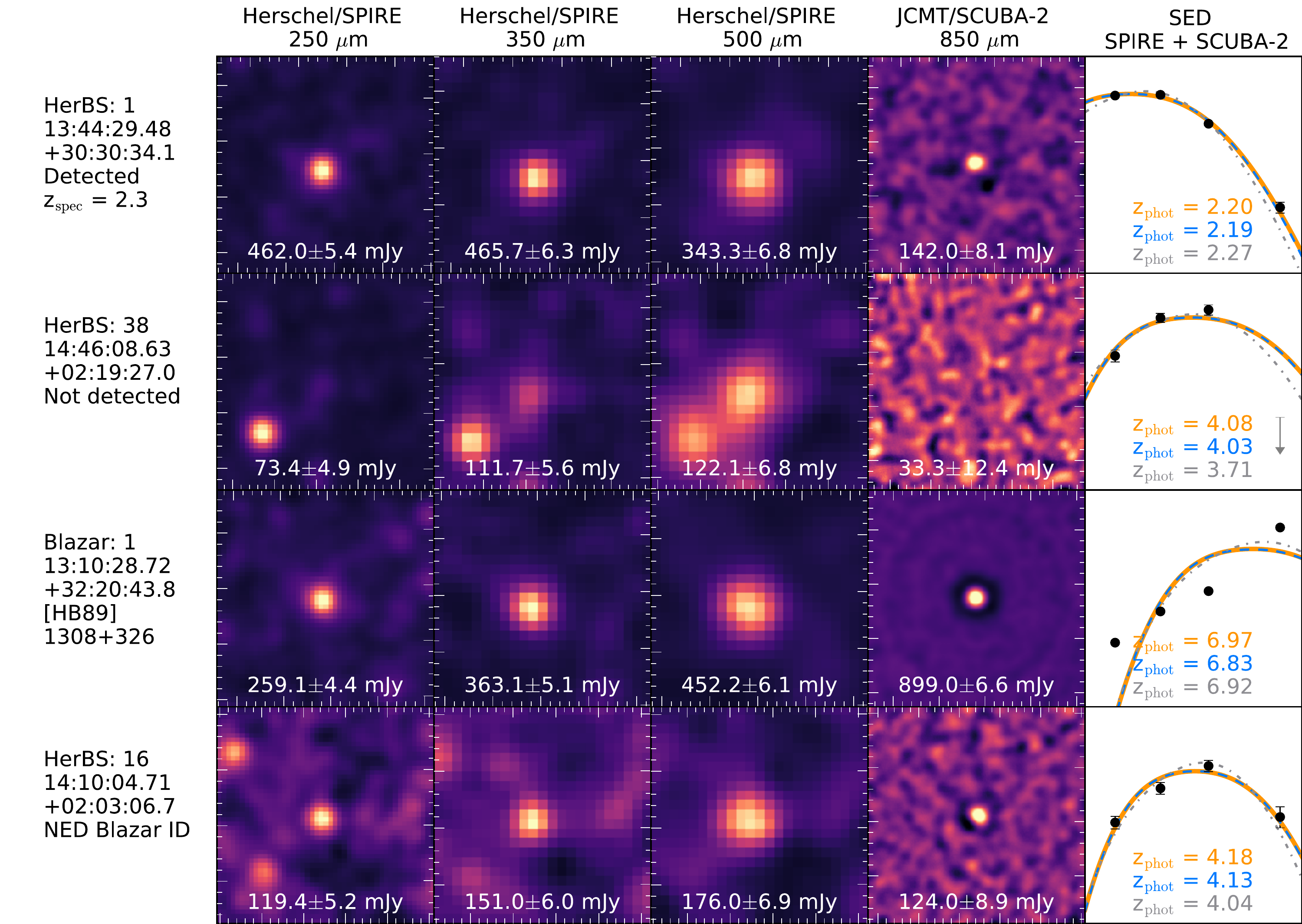}
    \caption{This figure shows the four different types of sources we found in the SCUBA-2 850 $\mu$m observations of our sample: a galaxy detected with SCUBA-2, a galaxy undetected with SCUBA-2, a blazar, and HerBS-16, which is close to a known blazar, but has an SED typical of thermal emission from dust. The first three columns of cutouts of each source are the \textit{Herschel} observations shown in 4 by 4 arc minute poststamps. The fourth column shows the 850 $\mu$m SCUBA-2 observation in a 4 by 4 arc minute poststamp. \textcolor{referee}{All poststamps are centred at the 250 $\mu$m extraction position of the Herschel catalogue.} The final frame is a fitted SED, with the \textit{best-fit} template in orange, \textit{fixed $\beta$} template in blue and Pearson's template in grey \citep{Pearson2013}. Similar figures for the entire HerBS sample can be found in Appendix \ref{sec:AppC}.}
    \label{fig:subsample}
\end{figure*}

One of the sources, HerBS-16, does not have the typical flux ratios of a blazar, and has therefore not been removed. The spectrum also looks dust-like, and has consistent photometric redshift estimates, as can be seen in Figure \ref{fig:subsample}. The source, in this case, could be close to the blazar by accident. Only one source close to a known blazar has not been observed, and we have therefore kept it in our HerBS sample (HerBS-112).

The difference between the graphs indicates the need for multi-wavelength observations, in order to reliably remove blazar contaminants from the sample. We list the \textit{Herschel} SPIRE and SCUBA-2 positions and fluxes of the removed blazars in Table \ref{tab:AppendixBlazars}.

After removing fourteen blazars from our sample, we are left with 189 HerBS galaxies with SCUBA-2 observations. While some sources close to NED blazars did not have irregular flux ratios, all of the sources with irregular flux ratios are close to known blazars. This suggests our method for finding contaminants in our sample is robust, and thus that the 19 unobserved sources that do not lie close to a NED blazar are not likely to have emission dominated by synchrotron radiation.

For completeness, we plot the blazar spectrum, assuming solely synchrotron radiation, in Figure \ref{fig:500vs850}, following equation
\begin{equation}
	S_{\nu} = A \times{} \nu^{-\alpha}.
\end{equation}
Here S$_{\nu}$ is the flux density at a specific frequency ($\nu$), $A$ is a constant factor, and $\alpha$ determines the steepness of the slope in the far-infrared wavelength regime. Most of the blazars lie close to this line. We also calculate the value for $\alpha$ for each galaxy, by minimizing $\chi^2$:
\begin{equation}
	\chi^2 = \sum^{i > j} \left[\frac{(S_i/S_j)_{\rm{model}} - (S_i/S_j)_{\rm{meas}}}{\sigma_{i,j,meas}}\right]^2.
\end{equation}
The index $i$ and $j$ iterate over all four wavelengths (250, 350, 500 and 850 $\mu$m), where $i$'s wavelength is always larger than $j$. $\sigma_{i,j,meas}$ is the combined error of $(S_i/S_j)_{\rm{meas}}$. $\alpha$-values range from 0.24 to 1.66. The individual values can be found in Table \ref{tab:AppendixBlazars}, and agree well with the positions of the blazar sources in Figure \ref{fig:500vs850}.

We provide poststamp cutouts of the observations with SPIRE, SCUBA-2 and fits of our templates (Section \ref{sec:tempfit}) to the 250, 350, 500 and 850 $\mu$m flux densities of each source in Appendix \ref{sec:AppC}. Typical cutouts of a source detected by SCUBA-2, a source undetected by SCUBA-2, and a blazar are shown in Figure \ref{fig:subsample}. The bottom row of cutouts shows HerBS-16, which is close to a NED blazar, but has an SED typical of a sub-mm galaxy.

%
%\begin{table}
%	\centering
%	\caption{SCUBA-2 observation results and sample purity}
%	\label{tab:Observations}
%	\begin{tabular}{lcccc}
%	\hline
%		 	& S/N < 3 	& 3 < S/N < 5 & 5 < S/N & Total \\
%		 	& \multicolumn{3}{c}{$\theta$ < 10" ( $\theta$ > 10")} & \\
%		\hline
%		\textbf{Original catalogue} 		& - & - & - & 223 \\
%		\textbf{SCUBA-2 obs.}			&	7 (20)	&	60 (15)		&	99 (2) & 203 \\
%		\textbf{Unobserved}  			& - & - & - & 20  \\ 
%		\textbf{Blazars}				&	0 (0)	&	0 (0)		&	14 (0) & 14 \\ \hline
%		\textbf{HerBS (SCUBA2)} 		& 27  & 75 	& 85       & 189  \\
%		\textbf{HerBS detected} 		& 0   & 60 	& 85       & 145  \\
%		\textbf{HerBS total} 	    		& 27  & 75 	& 87       & 209  \\
%	\end{tabular}
%	%\\ \flushleft \vspace{0.2cm} \textbf{Notes:} Reading from the left, the columns are: Column 1 - name of field; Column 2 and 3 - The location of the centre of the field; Column 4 and 5 - The approximate dimensions of the field; Column 6 - The surface area from the final maps \citep{Valiante2016}; Column 7 - The number of final HerBS sources in each field; Column 8 - The surface density of HerBS sources per field.
%\end{table}
\begin{table}
	\centering
	\caption{SCUBA-2 observations of the HerBS sample}
	\label{tab:Observations}
	\begin{tabular}{llrr}
		\hline
		& & Sources & Percentage\\
		\hline
		\multicolumn{2}{l}{\textbf{HerBS galaxies}}		& \textbf{209}	& \textbf{100 \%} 		\\
		\hline
		\multicolumn{2}{l}{\textbf{SCUBA-2 observed}}	& \textbf{189} 	& \textbf{90.4 \%}\vspace{0.0cm}\\
		Detected & ($>3\sigma$, $\theta$ < 10")			& 152 			&  69.4 \% 				\\
		Not detected & ($<3\sigma$)						& 27 			&  12.9 \% 				\\
		Not detected & ($>3\sigma$, $\theta$ > 10")		& 10 			&  8.1 \%\vspace{0.1cm}\\
		\multicolumn{2}{l}{\textbf{Not observed}}		& \textbf{20}	&  \textbf{9.6 \%} 		\\
		\hline
		\multicolumn{2}{l}{\textbf{Blazar contaminants}}		& 14 	&  
	\end{tabular}
\end{table}

\begin{table}
	\centering
	\caption{Re-examined SCUBA-2 observations of HerBS sources with $\theta$ $>$ 10 arc second.}%The maximum flux of these HerBS sources lay outside of a ten arcsecond circle from the SPIRE position. Rerunning flux extraction within a ten arcsecond circle resulted in seven extra detected sources.
	\label{tab:new850s}
	\begin{tabular}{lrrr}
		\hline
		HerBS 			& $\theta$ 		& S/N 			& S$_{\rm{850 \mu m}}$ \\
		$\rm{[-]}$ 		& $\rm{["]}$ 	& $\rm{[-]}$ 	& $\rm{[mJy]}$ \\
		\hline
		\textbf{63}		& \textbf{9.45} & \textbf{3.19} & \textbf{33.8} \\
		\textbf{75} 	& \textbf{7.59} & \textbf{4.24} & \textbf{44.9} \\
		96 				& 7.84 			& 2.10 			& 19.5 \\
		97 				& 6.57 			& 2.49 			& 28.1 \\ \vspace{0.15cm} 
		\textbf{101} 	& \textbf{1.93} & \textbf{3.42} & \textbf{32.5} \\
		118 			& 2.28 			& 2.12 			& 23.3 \\
		122 			& 6.97 			& 2.43 			& 21.9 \\
		131 			& 5.54 			& 2.95 			& 30.3 \\ 
		\textbf{140} 	& \textbf{7.14} & \textbf{3.59} & \textbf{30.3} \\ \vspace{0.15cm}
		\textbf{145} 	& \textbf{9.59} & \textbf{3.17} & \textbf{33.0} \\
		146 			& 7.85 			& 2.92 			& 32.1 \\
		\textbf{148} 	& \textbf{5.40} & \textbf{3.02} & \textbf{29.0} \\
		151 			& 6.33 			& 2.34 			& 23.9 \\ 
		163 			& 6.66 			& 1.85 			& 19.1 \\ \vspace{0.15cm}
		172 			& 5.92 			& 1.40 			& 13.7 \\
		\textbf{181} 	& \textbf{4.06} & \textbf{3.81} & \textbf{32.9} \\
		195 			& 3.94 			& 2.61 			& 29.5 \\
		
	\end{tabular}
\end{table}

% ++++++ Data Analysis ++++++
\section{Galaxy templates}
\label{sec:data}
\label{sec:template}
We derived a galaxy template for our total sample, by using the subset of HerBS sources that have spectroscopic redshifts. We fitted a two-temperature, modified blackbody spectral energy distribution to the \textit{Herschel} and the SCUBA-2 flux densities of each source. We list the sources with spectroscopic redshifts in Table \ref{tab:specsources}. These spectroscopic redshifts were found by observing sub-mm spectral lines, in order to ensure we are looking at the same source.

This template is necessary to estimate photometric redshifts and luminosities for our entire sample. Similar to the analysis of \cite{Pearson2013}, we fitted the template to the SPIRE (250, 350, and 500 $\mu$m) fluxes, and included our JCMT/SCUBA-2 850 $\mu$m flux densities. We choose to exclude the PACS photometry of our sources in our analysis, as even the brightest sources are poorly detected, due to the high-redshift limit of our sample. Our spectroscopic sample includes 8 sources used in Pearson's analysis, and 16 new sources, all of which are at high redshifts (z$_{\rm{spec}} > 1.5$). We only used HerBS sources for our template to ensure there is 850 $\mu$m photometry of our sources, and only used the galaxies with spectroscopic redshifts estimated from more than one line.

\begin{table}
	\centering
	\caption{The sources from the HerBS sample with measured spectroscopic redshifts.}
	\label{tab:specsources}
	\begin{tabular}{llcccc}
\hline
\multicolumn{6}{c}{\textbf{Robust, multi-line detections}}\\
\hline
	H-ATLAS name: & HerBS & z$_{\rm{spec}}$ & z$_{\rm{phot}}$ & $\Delta$z/(1+z) & Ref. \\
	\hline
J083518.4+303034 & 1 & 2.30 & 2.20 & 0.03 & H12 \\
J114637.9-001132 & 2 & 3.26 & 2.80 & 0.11 & H12 \\
J082403.8+334407 & 3 & 2.95 & 3.75 & -0.20 & H-p \\
J083051.0+013225 & 4 & 3.63 & 3.09 & 0.12 & R-p \\
J080520.2+233627 & 5 & 3.57 & 3.72 & -0.03 & R-p\vspace{0.15cm}\\
J082246.8+284449 & 6 & 1.68 & 2.11 & -0.16 & G13 \\
J082537.0+292326 & 7 & 2.78 & 2.89 & -0.03 & K-p \\
J084933.4+021442 & 8 & 2.41 & 2.64 & -0.07 & L-p \\% Studied by Ivison
J080214.5+261457 & 9 & 3.68 & 3.87 & -0.04 & K-p \\
J113526.2-014606 & 10 & 3.13 & 2.32 & 0.20 & H12\vspace{0.15cm}  \\
J082620.3+245900 & 12 & 3.11 & 2.29 & 0.20 & R-p \\
J142413.9+022303 & 13 & 4.28 & 4.53 & -0.05 & C11 \\
J141351.9-000026 & 15 & 2.48 & 2.62 & -0.04 & H12 \\
J090311.6+003907 & 19 & 3.04 & 3.76 & -0.18 & F11 \\% SDP 81
J082310.2+311534 & 20 & 1.84 & 1.88 & -0.02 & R-p \vspace{0.15cm} \\
J083144.0+255054 & 29 & 2.34 & 2.69 & -0.11 & R-p \\
J082153.5+341649 & 30 & 2.19 & 3.28 & -0.34 & R-p \\
J091840.8+023048 & 32 & 2.58 & 3.03 & -0.13 & H12 \\
J082949.3+300401 & 35 & 2.68 & 2.73 & -0.01 & H-p \\
J091304.9-005344 & 59 & 2.63 & 2.87 & -0.07 & N10 \vspace{0.15cm} \\
J115820.1-013752 & 66 & 2.19 & 2.49 & -0.09 & H-p \\
J113243.0-005108 & 71 & 2.58 & 3.73 & -0.32 & R-p\\
\hline
\multicolumn{6}{c}{\textbf{Tentative, single line detections (not used)}}\\
\hline
J080532.7+275900 & 31 & \textit{2.79} & 3.25 & -0.12 & - \\
J083344.9+000109 & 88 & \textit{3.10} & 3.25 & -0.04 & - \\
J113803.6-011737 & 96 & \textit{3.15} & 2.88 & -0.07 & H12 \\
J113833.3+004909 & 100 & \textit{2.22} & 2.66 & -0.14 & -
\end{tabular}
\\ \vspace{0.2cm} \raggedright \textbf{Notes:} Reading from the left, the columns are: Column 1 - the official H-ATLAS name; Column 2 - HerBS number; Column 3 - spectroscopic redshift; Column 4 - photometric redshift using the \textit{best-fit} model; Column 5 - $(z_{\rm{spec}} - z_{\rm{phot}})/(1+z_{\rm{spec}})$; Column 6 - Reference for the spectroscopic redshift: N10 $=$ \citet{Negrello2010}, F11 $=$ \citet{Frayer2011}, H12 $=$ \citet{Harris2012}, G13 $=$ \citet{George2013}, L13 $=$ \citet{Lupu2012}, B13 $=$ \citet{Bussmann2013}, H-p $=$ \citet{Harris-p}, R-p $=$ \citet{Riechers-p}, K-p $=$ \citet{Krips-p}, L-p $=$ \citet{Lupu-p}.
\end{table}

\subsection{Template fitting}
\label{sec:tempfit}
We fitted the template to the sources' flux densities and rest wavelengths, calculated from their spectroscopic redshifts. We assumed a two-temperature modified blackbody template for the SED,
\begin{equation}
S_{\nu} = A_{\rm{off}}\left[ B_{\nu} \left(T_h\right) \nu^{\beta} + \alpha B_{\nu} \left( T_c \right) \nu^{\beta} \right],
\label{eq:BB}
\end{equation}
where $S_{\nu}$ is the flux at the rest-frame frequency $\nu$, $A_{\rm{off}}$ is the normalisation factor, $B_{\nu}$ is the Planck blackbody function, $\beta$ is the dust emissivity index, $T_h$ and $T_c$ are the temperatures of the hot and cold dust components, and $\alpha$ is the ratio of the mass of the cold to hot dust. 

We aimed to minimize the following $\chi^2$ for the fluxes that were detected, 
\begin{equation}
\chi^2 = \sum^n_{i = 1}  \chi_i^2= \sum^n_{i = 1} \sum^{\lambda} \left[ \frac{A_iS_{\text{model},i} - S_{\text{meas},i}}{\sigma_{\text{meas},i}}\right]^2,
\label{eq:chi}
\end{equation}
where $S_{\text{model},i}$ is the predicted flux of the $i^{\text{th}}$ source (out of $n$) according to equation \ref{eq:BB}, with the amplitude $A_{\rm{off}}$ set to one. $S_{\text{meas},i}$ and $\sigma_{\text{meas},i}$ are the measured signal and noise values. In the case all fluxes of the source were detected, we fitted the amplitude of our template, $A_i$, to the rest-wavelength data points analytically in order to decrease computation time,
\begin{equation}
A_i = \left(\sum^{\lambda} \frac{S_{\text{model},j} S_{\text{meas},j}}{\sigma_{\text{meas},j}^2}\right) \Biggm/ \left(\sum^{\lambda} \frac{S_{\text{model},j}^2}{\sigma_{\text{meas},j}^2}\right).
\label{eq:A}
\end{equation}
Equation \ref{eq:A} is derived by solving d$\chi_i^2/$d$A_i$ = 0. We left the 

One source with a spectroscopic redshift did not have a detected SCUBA-2 flux, HerBS-71. In this upper-limit case, we calculated the $\chi^2$ contribution using the method detailed in \cite{Sawicki2012} and \cite{Thomson2017},
\begin{equation}
	\chi^2 = - 2 \sum_j \ln \int_{-\infty}^{3\sigma} \exp \left[-\frac{1}{2} \left(\frac{f - A_jS_{model,j}}{\sigma_{meas,j}}\right)^2 \right] df,
\end{equation}
where we sum over all non-detections $j$, which in our case is only the SCUBA-2 flux of HerBS-71, and integrate the gaussian distribution up to the detection criterion of three times the measured noise (3$\sigma$). 
The modified $\chi^2$ statistic quantifies the probability of an event where the noise affected the signal to drop below the detection criterion. In the case the model predicts a flux under the detection limit,  there is no discrepancy with the model, and we set the $\chi^2$-value to zero.

We did this template fitting for two templates: \textit{best-fit}, where we varied all the parameters ($T_c$, $T_h$, $\alpha$, and $\beta$), and \textit{fixed $\beta$} where we varied all parameters except $\beta$, which we fixed to 2.
We also tried keeping $T_{{c}}$, $T_{{h}}$, $\alpha$ and $\beta$ fixed to the values found by \cite{Pearson2013}. In this case we found the set of $A_{\rm{i}}$ that gave the minimum $\chi^2$ fit. The point of this was to determine whether our new templates gave any improvement in the quality of fit over that found by \cite{Pearson2013}.
We estimated the uncertainty on each parameter by incrementally changing this parameter until the minimised $\chi^2$ changes by of one (one interesting parameter, \citealt{Avni}). The $\chi^2$ was minimised by allowing the other (two or three) parameters to vary. The best-fit templates are given in Table \ref{tab:specsources}.

%We also use the same fitting method on subsamples of our sources with a spectroscopic redshift, in the hope to tease out some physical differences between these galaxies in our sample. We note that this method does run the risk of probing only specific regions of the template.
%We check the effect of redshift, by creating two subsamples split at z$_{\rm{spec}}$ $=$ 2.65. Similarly, we check the effect of observed luminosity, by creating two subsamples split at L$_{\rm{obs}}$ $=$ 5.1 $\times$ 10$^{13}$ L$_{\odot}$. Each subsample consists of twelve sources with a spectroscopic redshift.

\begin{figure}
	\includegraphics[width=\columnwidth]{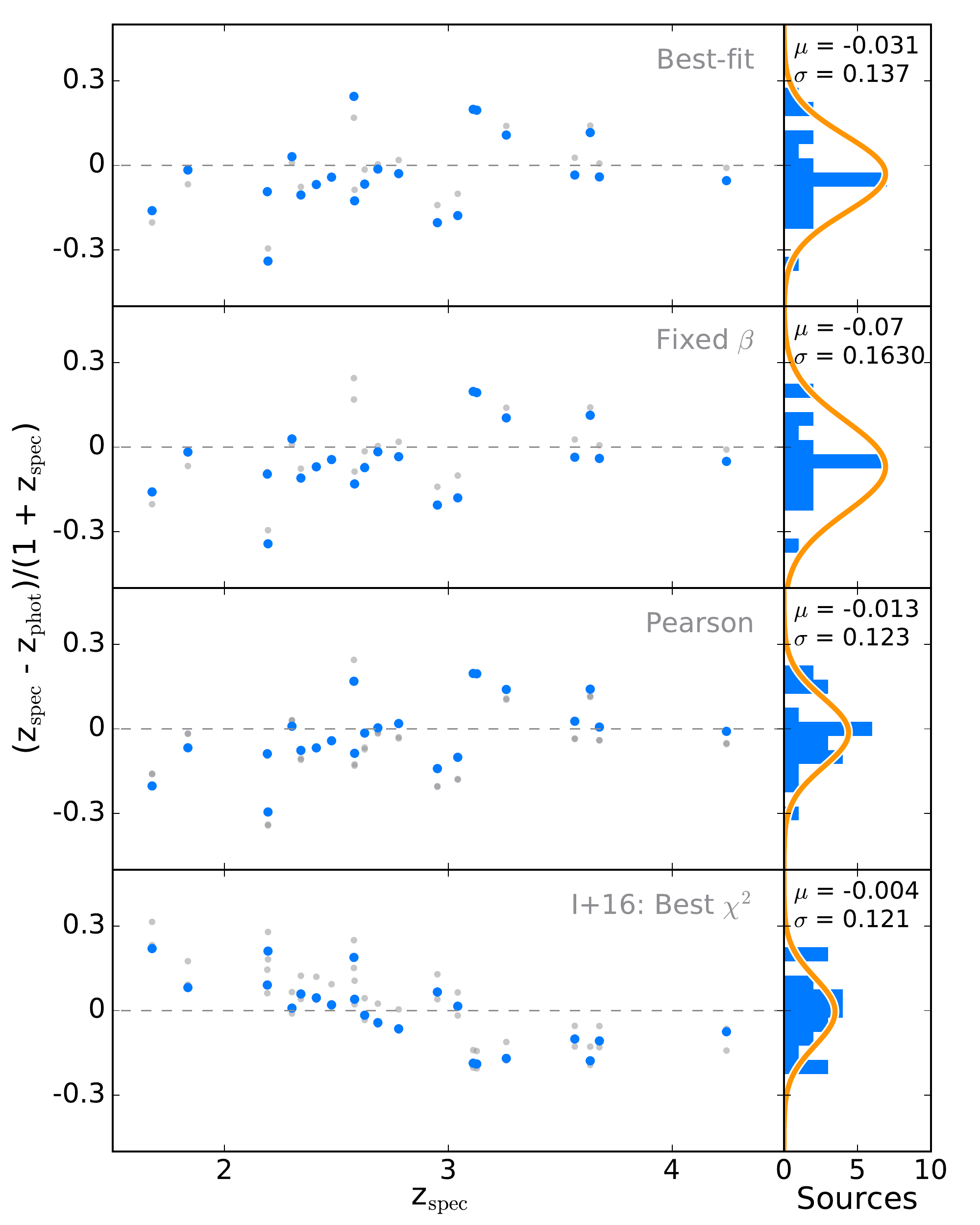}
    \caption{The top three panels show $(z_{\rm{spec}} - z_{\rm{phot}})/(1+z_{\rm{spec}})$ plotted against the spectroscopic redshift for the three templates. The \textit{blue dots} in each panel show the points for the specified template, while the \textit{smaller grey dots} show the points for the other two templates. The bottom panel shows $(z_{\rm{spec}} - z_{\rm{phot}})/(1+z_{\rm{spec}})$ for the three templates used for the redshift estimation in \citet{Ivison2015}, where the \textit{blue dots} correspond to the template fit with the lowest $\chi^2$ for each source individually, and the \textit{smaller grey dots} are the values of the two remaining templates.} 
    \label{fig:dz}
\end{figure}

\subsection{Template results}
We find a cold- and hot-dust temperature of  $21.29_{-1.66}^{+1.35}$ K and $45.80_{-3.48}^{+2.88}$ K, a cold-to-hot dust mass ratio of $26.62_{-6.74}^{+5.61}$ and a $\beta$ of $1.83_{-0.28}^{+0.14}$ for the \textit{best-fit} template. The results for the other templates, including the fitting of the templates to redshift and luminosity subsets, can be found in Table \ref{tab:subfitting}. 

We investigated the usefulness of each template for estimating photometric redshifts, by using each template to estimate the photometric redshift of each source, and then calculating $(z_{\rm{spec}} - z_{\rm{phot}})/(1+z_{\rm{spec}})$ for each source. 
The root mean squared value of $(z_{\rm{spec}} - z_{\rm{phot}})/(1+z_{\rm{spec}})$ for the \textit{best-fit} template is 13 \%, which is similar to the \textit{fixed-$\beta$} and Pearson templates.
The value of the relative error derived from the \textit{best-fit} template for each source is given in Table \ref{tab:specsources}, and the mean and standard deviations of this quantity for each template are given in Table \ref{tab:subfitting}. 

Figure \ref{fig:dz} shows $(z_{\rm{spec}} - z_{\rm{phot}})/(1+z_{\rm{spec}})$ plotted against spectroscopic redshift for the three templates. The three distributions are very similar.
We compare the redshift estimates against the method used in \cite{Ivison2015}. They fit three different templates (ALESS \citep{Swinbank2014}, Cosmic Eyelash \citep{Ivison2010,Swinbank2010}, and the template from \cite{Pope2008}) to the flux measurements, and use the redshift estimate from the spectrum with lowest $\chi^2$-value. When we apply this method to our sample of sources with spectroscopic redshifts, we achieve a slightly better redshift accuracy of $\sim$12 \%.

We note that the uncertainty in photometric redshift estimation using our new template, obtained from SCUBA-2 and \textit{Herschel} measurements, is not actually any smaller than that using the template that \cite{Pearson2013} obtained from \textit{Herschel} measurements alone. We discuss the significance of this in the Section \ref{sec:discussion}.

\begin{figure}
	\includegraphics[width=\columnwidth]{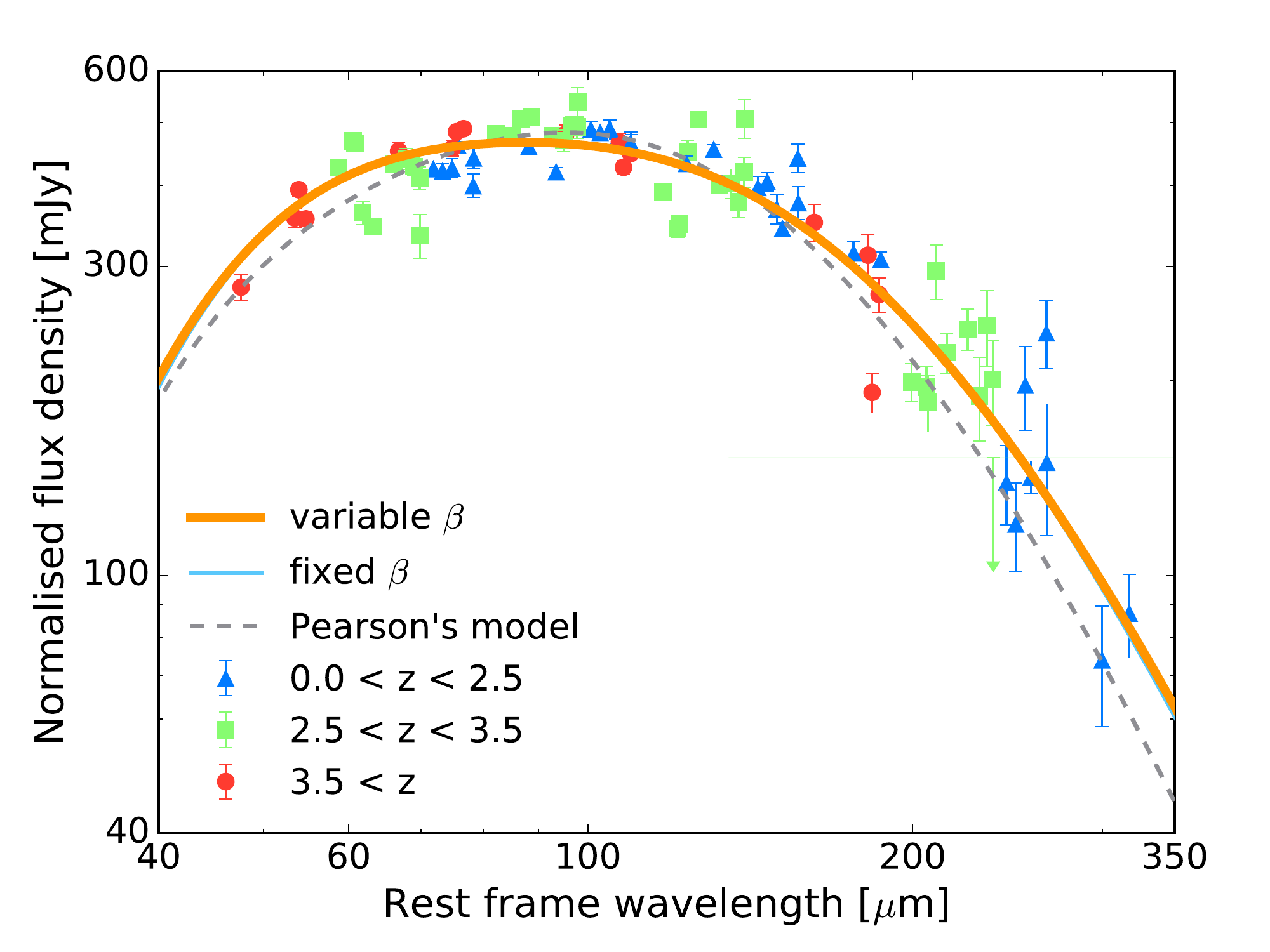}
    \caption{The flux densities of the spectroscopic sources plotted against rest-frame wavelength. The curves show the three templates (\textit{best-fit} is the \textit{thick orange line}, \textit{fixed-$\beta$} is the \textit{thin blue line}, and \textit{Pearson's model} is the \textit{dashed grey line}), and all the flux densities of each source are scaled to produce the same bolometric luminosity as the brightest source (HerBS: 1). The sample is split up in three redshift intervals, to associate each galaxy's four data points more easily.}
    \label{fig:fitting}
\end{figure}

Figure \ref{fig:fitting} shows the normalised flux densities of the spectroscopic sources against their rest-frame wavelength, with the three templates overlaid. The flux-densities are normalised to give each galaxy the same bolometric luminosity as HerBS-1. 

%Figure \ref{fig:subfitting} shows similar plots for the four templates obtained from the redshift and luminosity sub-samples. The flux-densities are normalised to give each galaxy the same bolometric luminosity as HerBS-1.

We used the photometric redshifts estimates of our \textit{best-fit} template to derive observed bolometric luminosities of the HerBS sources. As the redshift estimates are determined from a different spectrum, some of the photometric redshift estimates, z$_{\rm{phot}}$, fall below two. They are, however, kept in the HerBS sample, as not to increase the complexity of the selection functions. 

\begin{figure}
	\includegraphics[width=\columnwidth]{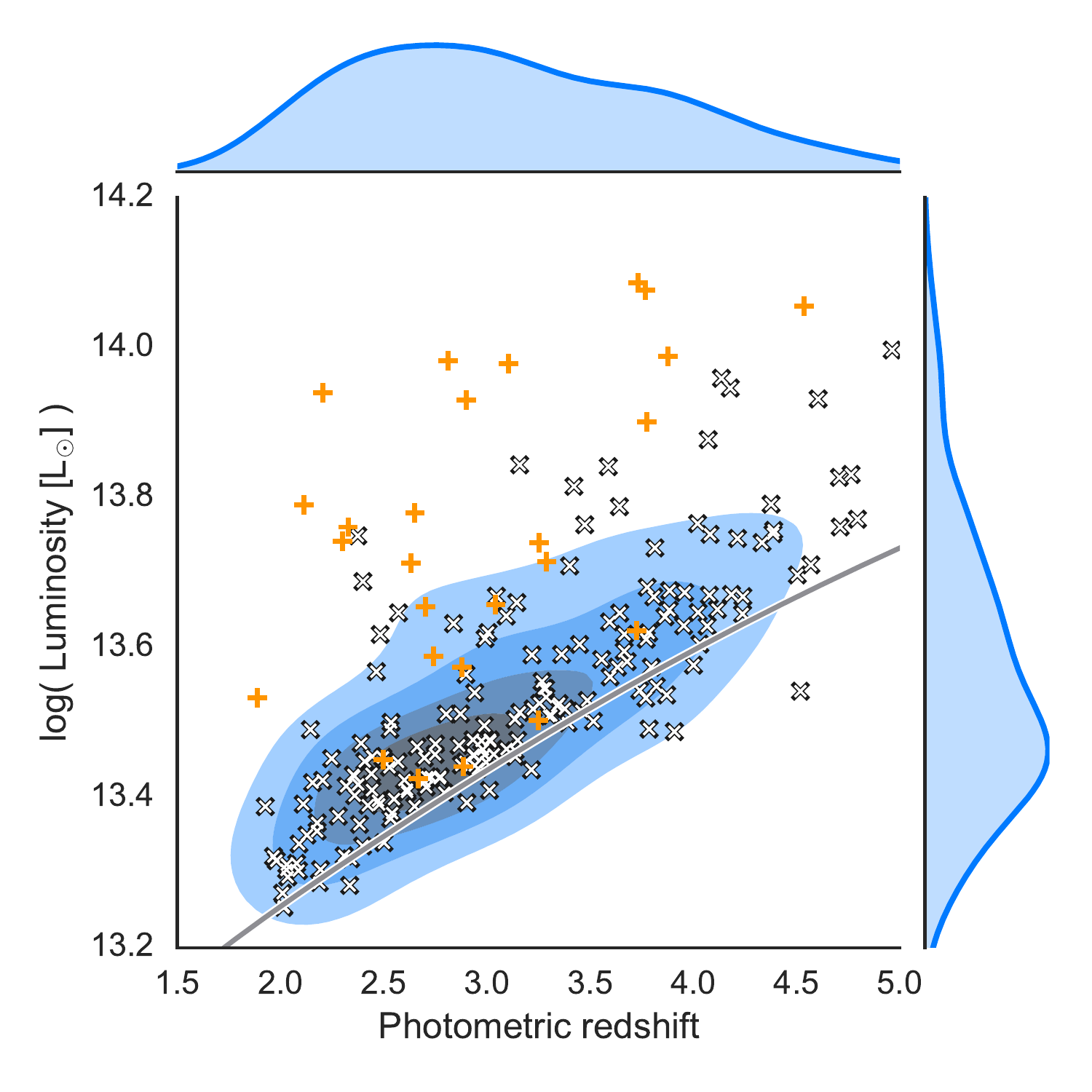}
    \caption{Observed bolometric far-infrared luminosity ($\lambda_{\rm{rest}} =$ 8 - 1000 $\mu$m) plotted against photometric redshift, calculated with the \textit{best-fit} template. Sources with spectroscopic redshifts are plotted in \textit{orange plusses}, although the redshifts used in the diagram are their photometric redshifts. The smoothed distributions of redshift and luminosity are shown on the sides of the scatter plots. The grey line shows bolometric luminosity for the \textit{best-fit} template, assuming S$_{\rm{500\mu m}}$ = 80 mJy, as a function of redshift.}
    \label{fig:hist}
\end{figure}

We calculate the observed bolometric luminosities by deriving the photometric redshift from our \textit{best-fit} template, and integrating the template from $\lambda_{\rm{rest}} =$ 8 to 1000 $\mu$m. The estimated redshifts and bolometric luminosities are listed in Table \ref{tab:AppendixA}, as well as the photometric redshift estimates using the method from \cite{Ivison2015}. Figure \ref{fig:hist} shows the distribution of sources as a function of redshift and luminosity. This figure shows that the majority of our sources with a spectroscopic redshift are in the higher luminosity range, as typically spectroscopic campaigns aim for the brightest sources first.

\begin{table}
	\centering
	\caption{The results of the fitting of the total sample, with a variable and fixed beta, and applying the template from \citet{Pearson2013} to our sources.}
	\label{tab:subfitting}
	\begin{tabular}{lccc}
\hline
			& Total 					& Fixed-Beta 				& Pearson \\
\hline \vspace{0.1cm}
$T_c$ [K] 	& $21.29_{-1.66}^{+1.35}$ 	& $20.47_{-0.26}^{+0.26}$	& 23.9\\ \vspace{0.1cm}
$T_h$ [K] 	& $45.80_{-3.48}^{+2.88}$	& $44.05_{-0.55}^{+0.52}$	& 46.9\\\vspace{0.1cm}
$\alpha$ 	& $26.69_{-6.74}^{+5.61}$	& $30.46_{-1.42}^{+1.32}$	& 30.1\\\vspace{0.1cm}
$\beta$ 	& $1.83_{-0.28}^{+0.14}$		& 2	(fixed)					 & 2 (fixed)\\
$\chi^2$	& 812.58  					& 812.96						& 1101.03 \\
$\Delta$z/(z$_{\rm{spec}}$ + 1) 			& -0.03$\pm$0.14		& -0.03$\pm$0.14			& -0.01$\pm$0.12
\end{tabular}
\end{table}

\section{Discussion}
\label{sec:discussion}
\subsection{Source confusion}
\label{sec:SC}
We have selected our HerBS sample using a 500 $\mu$m flux limit. 
The large beam-width at this wavelength could cause us to confuse multiple line-of-sight sources into a single observed source, and hence yield a 500 $\mu$m flux density that is too large.

Observationally, high resolution studies of sub-millimetre galaxies show this to be the case, although the severity of this effect varies from study to study \citep{Hodge2013,Koprowski2014}.
An SMA study by \cite{Chen2013} of sources selected at 450 $\mu$m only found 10 \% of the sources to be significantly amplified by line-of-sight sources.
An ALMA survey of 870 $\mu$m selected ALESS sources finds that up to 50 \% of the sources are significantly affected \citep{Hodge2013, Karim2013}.
Longer wavelengths and higher selection flux densities correlate with more source confusion, although all observational multiplicity studies so far focus on SMGs with a low probability of lensing.

A recent study by \cite{Scudder2016} used Bayesian inference methods to estimate the effects of source confusion in \textit{Herschel} observations at 250 $\mu$m.
They concluded that individual 250 $\mu$m sources are often the combination of emission from more than one galaxy. 

The solid angle of the beam of the JCMT at 850 $\mu$m is six times smaller than the beam of the 500 $\mu$m SPIRE observations.
We do not see any of our HerBS sources resolve into multiple $> 3\sigma$-detected components. 
This suggests that our long-wavelength observations are not confused, unless the sources are clustered on a scale smaller than the JCMT's beam size. The small clustering size could be the case, as \cite{Karim2013} finds the multiple emissions are separated less than 6" in the majority of cases of source confusion. Similarly, \cite{Chen2016} measured the clustering of SMGs on scales down to 1.5" using SCUBA-2 combined with deep near-infrared and optical data, and they also report a steep increase in angular correlation below 6". 
\textcolor{referee}{However, \cite{Hayward2013} simulated light cones to estimate the blending ratio of associated and unassociated SMGs for a 15 arcsecond beam, and found that at least 50 per cent of all blended SMGs show an unassociated SMG. The HerBS sources are selected by their 500 $\mu$m flux, which has a 36 arcsecond beam, and should therefore be more influenced by unassociated SMGs. As these unassociated SMGs are spatially unrelated to the source, they should have shown up in our JCMT analysis.}
A reason for the lack of source confusion could be due to our selection of lensed sources, as the probability for gravitational lensing is small, and two unrelated sources in the same \textit{Herschel} beam are unlikely to be both lensed by the same galaxy. 

Strong gravitational lensing could also be caused by a cluster of galaxies, which acts on a longer angular scale. These events are less common \citep{Negrello2016}, however \cite{Zavala2015} did report on the redshifts of cluster-lensed sources, one of which turned out to be three sources that was blended and lensed. We did not exclude these possibilities, however considering their infrequency, we can state that this lensing type would not influence the entire sample.

\subsection{The diversity of galaxies}
\label{sec:div}
In Section \ref{sec:template}, we fitted a two-temperature modified blackbody template to 22 HerBS sources with spectroscopic redshifts, the results of which can be seen in Table \ref{tab:subfitting}.

Both the fixed-$\beta$ and \textit{best-fit} templates result in similar templates, as the $\beta$-value of the \textit{best-fit} template is similar within the error bars. The errors on the \textit{best-fit} template are slightly larger, as more parameters are being fitted.
The temperatures on both fitted templates are slightly cooler than the template from \cite{Pearson2013}, however we do not find an indication of a cool gas component with a temperature T < 20 K, as found in \cite{PlanckThermal} and \cite{Clements2010}. The values we find for the temperatures agree broadly with the initial fitting attempts by \cite{DunneEales2001}, and the overall findings of \cite{Clements2010}.

The large $\chi^2$ values in Table \ref{tab:subfitting} imply that a single template is not actually a good representation of the data.
We fit our template to 22 galaxies, each with 4 data points, except one source where we only fitted the three SPIRE fluxes, as its SCUBA-2 flux remained undetected.
The free parameters in our model are the template parameters (3 or 4) and the amplitudes for each galaxy (22, eq. \ref{eq:A}). The expected $\chi^2$ values for the two models, on the assumption that they are a good representation of the data, are therefore
\begin{eqnarray*}
\chi^2_{{Best-fit}} &\approx & N_{\rm{data}} - N_{\rm{param}} - 1 \\
&\approx & 4 \times{22} - 22 - 4 - 1 \\
&\approx & 61, \\
\chi^2_{{Fixed-\beta}} &\approx & N_{\rm{data}} - N_{\rm{param}} - 1 \\
&\approx & 4 \times{22} - 22 - 3 - 1 \\
&\approx & 62.
\end{eqnarray*}
However, we observe $\chi^2$-values of $\sim$812, indicating that our sources are poorly modelled by a single galaxy template.

%We have fitted a template to a subselection of the HerBS sources with a spectroscopic redshift, the results of which can be seen in Figure \ref{fig:subfitting}, and in Table \ref{tab:subfitting}. We either divided our sources in redshift or luminosity. Low-redshift galaxies give better constraints on the long-wavelength (cold) end of the template and vice versa for the high redshift galaxies. As can be seen in Figure \ref{fig:hist}, the sources with a low observed luminosity are mostly at low-redshift, which therefore also probes the cold part of the spectrum more than the high-luminosity galaxies. Especially the low luminosity plot seems poorly constrained on the short wavelength scale.

%Table \ref{tab:subfitting} shows some significant differences between the templates derived by dividing the samples by redshift of luminosity, but because the sub-samples are sensitive to different ranges of the rest-frame wavelength, we cannot be sure the different templates are revealing any physical differences between the SEDs. 

%Cautiously though, we could state that, especially for the \textit{fixed-$\beta$}, the templates look remarkably similar between all sources. The high spikes in $\alpha$ can be explained because the total flux emitted by a black-body scales to the fourth power, and hence a slightly higher or lower temperature influences the total flux dramatically.

We tested the photometric redshift estimates of the templates using the same sources we used to derive the \textit{best-fit} template. 
However, we found no improvement in accuracy (Table \ref{tab:subfitting}) compared to the older template of \cite{Pearson2013}.
Similarly, Figure \ref{fig:dz} shows a similar pattern of redshift errors for all three templates. The redshift estimation by \cite{Ivison2015} might provide a slightly better estimation of the redshift, which are therefore added to the catalogue Table \ref{tab:AppendixA}.
The explanation for this lack of improvement is almost certainly the diversity of the population; the limit on the accuracy of photometric redshift estimates is not set by the accuracy of the average template but by the fact that galaxies have different spectral energy distributions.

\subsection{Redshift distribution of the HerBS sample}
\label{sec:KS}
\begin{figure}
	\includegraphics[width=\columnwidth]{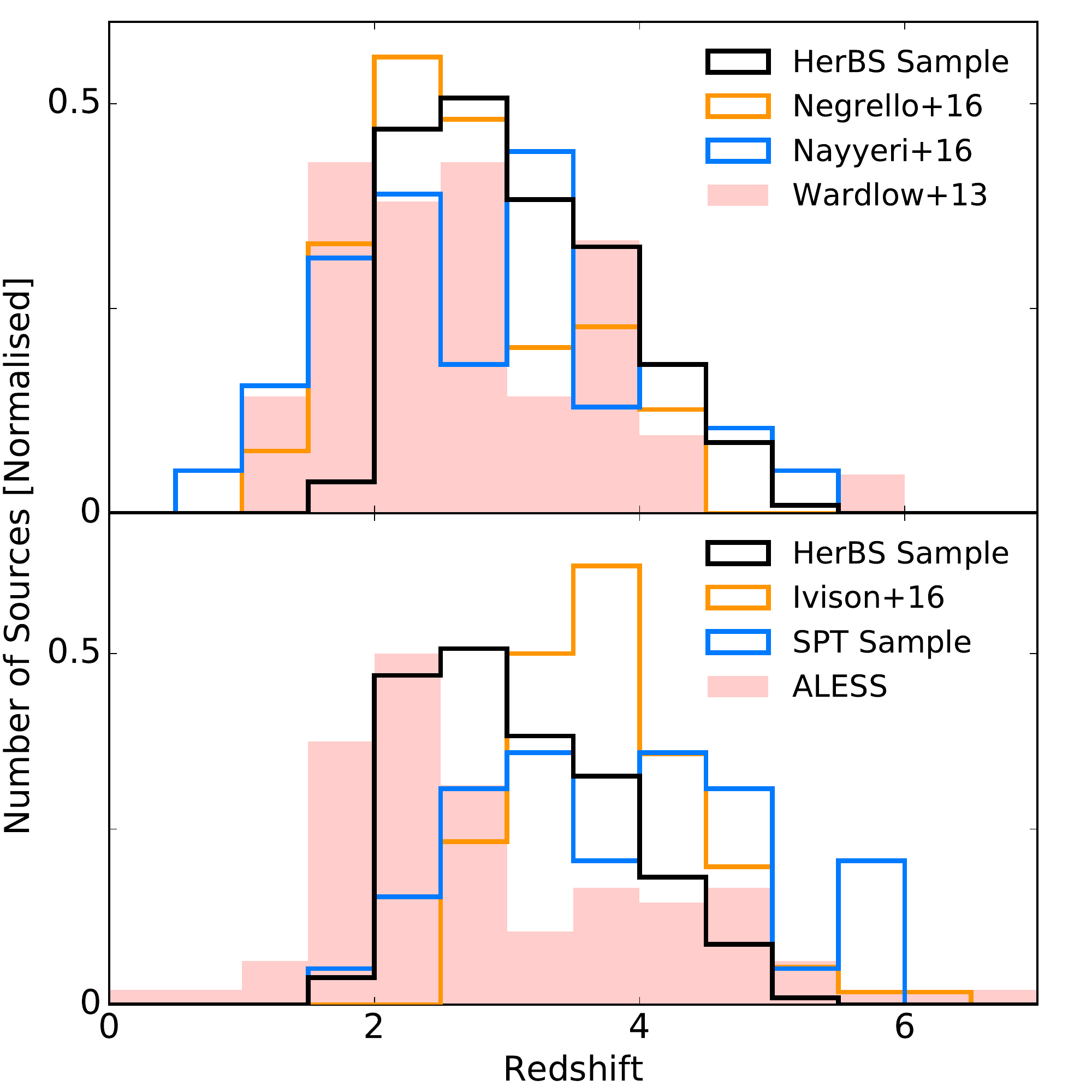}
    \caption{The top panel compares the redshift distribution of the HerBS sample (\textit{black}) to that of three samples selected with \textit{Herschel}/SPIRE at 500 $\mu$m.
    The bottom panel compares the redshift distribution of the HerBS sample (\textit{black}) to that of three samples with different selection wavelengths and colour cuts.}
    \label{fig:SPTcomp}
\end{figure}

Figure \ref{fig:SPTcomp} shows the redshift distribution of the HerBS sample, compared against various other galaxy samples, that are summarised in Table \ref{tab:zdist}. The top panel compares the distribution to samples selected with a simple flux cut-off at 500 $\mu$m. The sample from \cite{Negrello2016} used a S$_{\rm{500\mu m}}$ > 100 mJy flux cut on 600 sqr. deg. of the H-ATLAS field (they used a conservative mask on the SGP field). The sample from \cite{Nayyeri2016} used the same flux cut on the 372 sqr. deg. HeLMS and HeRS fields. We plot the total sample from \cite{Wardlow2013}. They used the 95 sqr. deg. HerMES survey, and their 500 $\mu$m flux cut-off went down to 80 mJy.

The bottom panel compares the HerBS redshift distribution against samples selected at various wavelengths. 
The sample from \cite{Ivison2015} is also from the H-ATLAS fields, and contains sources with a color-cut at S$_{\rm{500\mu m}}$/S$_{\rm{250\mu m}}$ > 1.5 and S$_{\rm{500\mu m}}$/S$_{\rm{350\mu m}}$ > 0.85, in order to select sources at high redshift. The sources were also selected to have relatively low 500 $\mu$m flux density of around 50 mJy, in order to select unlensed sources. Their unlensed nature reduces the uncertainty in the intrinsic luminosity of the source.
The South Pole Telescope (SPT) lensed sample was selected from 2500 sqr. deg. SPT survey by a flux cut at S$_{1.4\rm{mm}}$ $>$ 20 mJy, and demanding the source has a dust-like spectrum. Low-redshift sources were removed with radio and far-infrared flux limits \citep{Weiss2013,Strandet2016}. 
The ALESS sample is initially selected from the LESS sample at S$_{\rm{870\mu m}}$ > 4.4 mJy from the 0.25 sqr. deg. Extended Chandra Deep Field South (ECDFS) field \citep{Weiss2009}. ALMA observations of the LESS sample removed all contaminants, resulting in a final ALMA-LESS (ALESS) sample of 96 SMGs \citep{Simpson2014}.

All samples selected at 500 $\mu$m with a simple flux cut have a similar redshift profile, and do not differ significantly from the HerBS sample when we take the photometric redshift cut-off into account. Also, without the photometric redshift cut-off, the standard deviation of the HerBS sample would have been larger.

Typically, higher average redshifts are expected for longer selection wavelengths \citep{Bethermin2015}. We see this for the SPT sample, which has higher average redshifts. The ALESS sample, selected at 870 $\mu$m, has a higher average redshift than the 500 $\mu$m without redshift constraints, but a lower average redshift than the HerBS sample due to HerBS photometric redshift constraint. The SPT and ALESS samples have a larger standard deviation in their redshifts, because the K-correction is negative for wavelengths between 850 $\mu$m and $\sim$3 mm. Comparison with the Ivison sample is difficult because of the more complicated selection criteria they employ.

A way of quantifying the similarity between the samples is using the Kolmogorov-Smirnov test. We compare each sample's sources with a redshift (spectroscopically or photometrically determined) greater than 2 to the photometric redshifts of the HerBS sources with z$_{\rm{phot}}$ > 2. For each sample, we run this method 100.000 times while randomly varying the redshift of each source acccording to a gaussian distribution with a width of $\Delta$ z = 0.15(1 + z). For the comparison to Ivison's sample, we only compare it to HerBS sources with a similar colour cut as they employed (S$_{\rm{500\mu m}}$/S$_{\rm{250\mu m}}$ > 1.5 and S$_{\rm{500\mu m}}$/S$_{\rm{350\mu m}}$ > 0.85), which only 26 HerBS sources follow. For the SPT sample, we used our best-fit template to estimate the flux at 1.4mm, and only compared the sources that follow the SPT flux cut (S$_{1.4\rm{mm}}$ $>$ 20 mJy), a property only 60 HerBS sources have. The ALESS flux criterion (S$_{\rm{870\mu m}}$ > 4.4 mJy) was also estimated using the best-fit template, and was met by all our 209 sources.

We detail the KS probability values in terms of disagreement between two samples in standard deviations ($\sigma$) in Table \ref{tab:zdist}. A comparison between the redistributed redshifts and the original, unvaried redshift estimates of the HerBS sources gives a 1.27 $\pm$ 0.45 times the standard deviation, which indicates we should expect rather large uncertainties in the probability measurements. The spectroscopic redshifts of the HerBS sources disagree with 2.01 $\pm$ 0.31 times the standard deviation with the redistributed redshifts. When we compare the photometric redshift estimates of these spectroscopic sources to the HerBS sample, this value drops to 0.79 $\pm$ 0.56. Our HerBS sample thus appears probed evenly by the current set of HerBS sources with spectroscopic redshifts.

The sample from Negrello features more galaxies at low selected redshifts (2 < $z$ < 3), causing the disagreement seen by the relatively high KS value. This is contrary to both Nayyeri and Wardlow's samples, who agree strongly with the HerBS sample, suggesting that these sources are drawn from the same population. \textcolor{referee}{Only one out of four sources with low 500 $\mu$m flux densities ($\sim$80 mJy) in Wardlow's sample was found to be lensed. This seems contradictory to the high likeness with the HerBS sample, which has a high lensing fraction of 76 per cent, found in Section \ref{sec:lensingCalc}. Only four of Wardlow's sources were checked for their lensing nature, which could indicate that their low lensing fraction is caused by small-number statistics. We can also think of two physical reasons for the low lensing fractions, namely the absence of a redshift selection and the actual decrease in the lensed fraction at lower flux densities. Redshift selection lifts the probability of lensing, by ensuring the sources are drawn from the redshift space most lensed sources are in \citep{Strandet2016}. Similarly, at lower flux densities, the fraction of lensed sources decreases, as can be seen in Figure \ref{fig:numbercounts}. }

The SPT also seem to probe similar populations to the HerBS sources, further increasing our suspicion of a high lensing fraction in our sample. A slightly less strong agreement with the ALESS sample was found, which probes deeper on a smaller part of the sky. Interestingly, \cite{Strandet2016} reports a disagreement of around 2.4 standard deviations between the SPT and ALESS sample. The HerBS sample likeness to the SPT sample is larger, suggesting this sample is more similar than to the deeper ALESS sample, especially as \cite{Strandet2016} found those two samples to be different. This is further proven by the small lensing fraction in the ALESS sample, compared to the sizeable lensing fraction in the SPT sample, and the lensing fraction we find in Section \ref{sec:lensingCalc}. \textcolor{referee}{However, \cite{Hodge2013} and \cite{Karim2013}'s studies of the ALESS sample did suggest a source confusion fraction on the order of 50 per cent of their sample. Even though our samples are not completely similar, this high blending percentage might indicate that our method of estimating the effects of source confusion with the JCMT's beam is incomplete.}
The low agreement to Ivison's sample suggests that their selection of unlensed SMGs was effective, and it indicates they might select different galaxies than our sample.

\begin{table*}
	\centering
	\caption{Redshift distributions of several sub-mm samples.}
	\label{tab:zdist}
	\begin{tabular}{lccccl} % four columns, alignment for each
		\hline
		Sample & $\langle$z$\rangle$ $\pm$ $\sigma$ & Sources  & Surface & KS $\sigma$-value & Selection criterion\\
		\hline
		HerBS 						& 3.09 $\pm$ 0.71 	& 209 & 616.4 & 1.27 $\pm$ 0.45 & S$_{\rm{500 \mu m}}$ > 80 mJy; z$_{\rm{phot}}$ > 2.0 \\
		HerBS with z$_{\rm{spec}}$ 	& 3.07 $\pm$ 0.72 	& 22  & 616.4 & 2.01 $\pm$ 0.31 & S$_{\rm{500 \mu m}}$ > 80 mJy; z$_{\rm{phot}}$ > 2.0	 \\ 
		\hline
		Negrello					& 2.64 $\pm$ 0.75	& 80 & 616.4& 1.82 $\pm$ 0.77 & S$_{\rm{500 \mu m}}$ > 100 mJy \\
		Nayyeri 					& 2.77 $\pm$ 1.02 	& 77 & 372 	& 0.66 $\pm$ 0.50 & S$_{\rm{500 \mu m}}$ > 100 mJy \\
		Wardlow 					& 2.65 $\pm$ 0.90 	& 42 & 95  	& 0.93 $\pm$ 0.66 & S$_{\rm{500 \mu m}}$ > 80 mJy	\\
		\hline
		Ivison 						& 3.80 $\pm$ 0.67	& 112 & 616.4 & 2.31 $\pm$ 0.84 & S$_{\rm{500 \mu m}}$ $\sim$ 50 mJy; S$_{\rm{500\mu m}}$/S$_{\rm{250\mu m}}$ > 1.5; S$_{\rm{500\mu m}}$/S$_{\rm{350\mu m}}$ > 0.85\\
		SPT sample					& 3.81 $\pm$ 1.07 	& 39  & 2500 	& 0.88 $\pm$ 0.55 & S$_{\rm{1.4 mm}}$ > 20 mJy		\\
		ALESS						& 2.90 $\pm$ 1.22	& 96  & 0.25 	& 1.26 $\pm$ 0.54 &  S$_{\rm{870 \mu m}}$ > 4.4 mJy
	\end{tabular}
\end{table*}

\subsection{Lensing fraction}
\label{sec:lensingCalc}
The SCUBA-2 observations do not resolve lensing directly, as the beam size (13") is much larger than the typical Einstein rings caused by galaxy-galaxy lensing ($\sim1$") \citep{Bussmann2013,ALMA2015}. However, we can estimate the lensing fraction of our sample when we compare the distribution of flux densities of our sources to the predictions of galaxy evolution models that include gravitational lensing.

Here we use the hybrid model by \cite{Cai2013} with a cut-off lensing magnification factor of $\mu =$ 30. The hybrid model is based on a parametric backward model for redshifts lower than 1.5, whilst it calculates galaxy evolution for redshifts greater than 1.0 using physical models for the evolution of proto-spheroidal galaxies and their associated AGN. The model matches these two approaches to each other in the region between redshift 1.0 and 1.5. We assume all unlensed sources are high-redshift, proto-spheroidal galaxies. We did not observe all of the sample at 850 $\mu$m, so we expect that our observed number counts are a lower limit. 
\begin{figure}
	\includegraphics[width=\columnwidth]{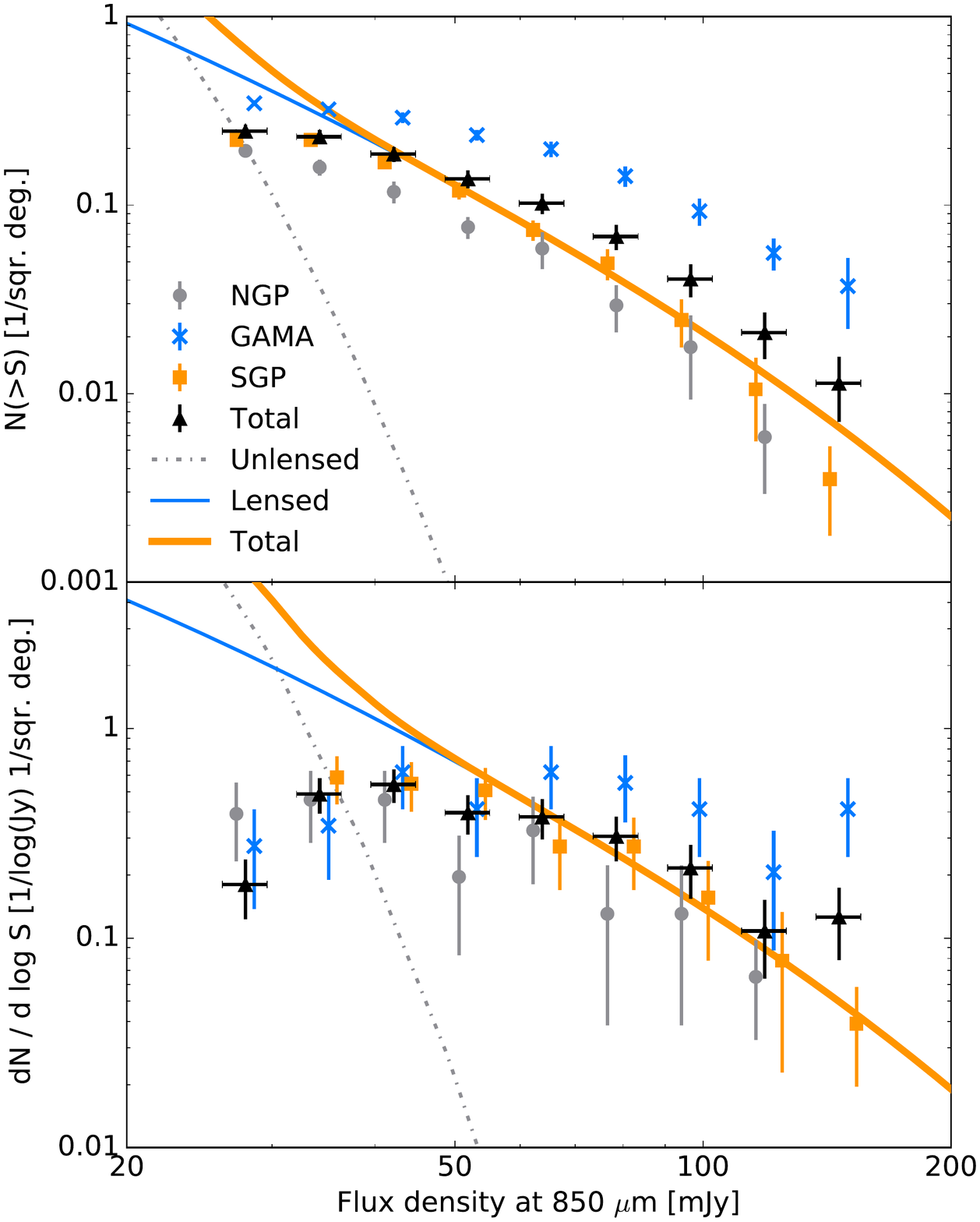}
    \caption{The top panel shows the cumulative number counts and the bottom panel shows the differential number counts of our HerBS sample, compared to the predictions of the model of \citet{Cai2013} for unlensed (\textit{dashed grey line}) and lensed (\textit{solid blue line}) galaxies.}
    \label{fig:numbercounts}
\end{figure}

Figure \ref{fig:numbercounts} shows a comparison of our number counts at 850 $\mu$m with the predictions of the model of \cite{Cai2013}. \textcolor{referee}{We have plotted the number counts for each of our fields, by summing the number of sources brighter than a given flux, and dividing by the corresponding area of the field, see Table \ref{tab:HATLAS}.} We estimate the error on the counts as the square root of the number of sources in each bin. A comparison of our counts with the predicted counts of the unlensed sources (\textit{grey dashed line}) immediately suggests most of our sources are lensed. We can quantify this as follows.

\textcolor{referee}{At the low fluxes, the data deviate from the model, because of the incompleteness of the HerBS sample at fluxes lower than $\sim$50 mJy. There are more sources than the model predicts at high fluxes, the significance of which is difficult to pin down due to the small number of sources. It is possible our sources have over-estimated 850 $\mu$m fluxes, possibly due to source confusion. However, it is important to realise that the model of \cite{Cai2013} is based on fitted luminosity functions. The high flux end of the luminosity function require large area surveys to be accurately fitted. As our sample is extracted from the largest area Herschel survey, the model is thus comparably uncertain as our data.}

We calculate the total number of lensed sources,
\begin{equation}
	\rm{N_{lens}(>S_{\nu})} = \sum_i^{\rm{N_{gal}}(> S_{\nu})} \rm{p_{lens}(S_{\nu,i})}.
	\label{eq:lensing}
	\end{equation}
We sum the lensing probability, $\rm{p_{lens}(S_{\nu,i})}$, over all galaxies brighter than the flux cutoff, $\rm{N_{gal}(> S_{\nu})}$. We calculate the probability, $\rm{p_{lens}(S_{\nu,i})}$, from the relative proportions of the differential number counts predicted for lensed and unlensed galaxies,
	\begin{equation}
\rm{p_{lens}(S_{\nu,i})} = \left[\frac{\rm{dN_{lens}}}{\rm{dS_{\nu}}}\left/ \left( \frac{\rm{dN_{proto}}}{\rm{dS_{\nu}}} + \frac{\rm{dN_{lens}}}{\rm{dS_{\nu}}} \right)\right.\right]_{\rm{S_{\nu,i}}}.
\end{equation}
The $\rm{N_{lens}}$ term refers to the lensed sources, and the $\rm{N_{proto}}$ term refers to the unlensed proto-spheroidal galaxies. We evaluate the probability at the flux density of the source, S$_{\rm{\nu,i}}$. Using the bottom panel of Figure \ref{fig:numbercounts}, $\rm{p_{lens}}$ can be thought of as the fraction lenses (\textit{thin blue line}) over the total sources (\textit{thick orange line}).
\begin{table}
	\centering
	\caption{Predicted lenses in the HerBS sample}
	\label{tab:lenses}
	\begin{tabular}{lrrr}
		\hline
		 \multicolumn{1}{l}{S$_{850 \rm{\mu m}}$ [mJy]} & \multicolumn{1}{r}{N(> S$_{850 \rm{\mu m}}$)} & \multicolumn{1}{r}{Lenses} & \multicolumn{1}{r}{Percentage} \\ \hline
all & 152.0 $\pm$ 0.0 & 128.4 $\pm$ 2.1 & 84.5 $\pm$ 1.4 \%\\
30 & 133.8 $\pm$ 3.4 & 123.3 $\pm$ 2.9 & 92.2 $\pm$ 0.9 \%\\
40 & 107.6 $\pm$ 3.9 & 105.2 $\pm$ 3.7 & 97.8 $\pm$ 0.3 \%\\
50 & 80.8 $\pm$ 3.6 & 80.5 $\pm$ 3.6 & 99.6 $\pm$ 0.1 \%\\
60 & 60.0 $\pm$ 3.2 & 59.9 $\pm$ 3.2 & 99.9 $\pm$ 0.0 \%\\
70 & 44.2 $\pm$ 2.9 & 44.2 $\pm$ 2.9 & 100.0 $\pm$ 0.0 \%\\
80 & 32.4 $\pm$ 2.4 & 32.4 $\pm$ 2.4 & 100.0 $\pm$ 0.0 \%\\
90 & 23.7 $\pm$ 2.0 & 23.7 $\pm$ 2.0 & 100.0 $\pm$ 0.0 \%\\
100 & 17.4 $\pm$ 1.7 & 17.4 $\pm$ 1.7 & 100.0 $\pm$ 0.0 \%\\
120 & 9.5 $\pm$ 1.3 & 9.5 $\pm$ 1.3 & 100.0 $\pm$ 0.0 \%
	\end{tabular}
\end{table}

We iterate this procedure a 1000 times, varying the 850 $\mu$m flux with a gaussian distribution with a width of the measurement uncertainty. Table \ref{tab:lenses} shows the predicted number of lensed sources (eq. \ref{eq:lensing}) and the observed number of sources for all SCUBA-2 detected HerBS sources. All of the errors are the standard deviations. Even for sources at S$_{\rm{850 \mu m}}$ $>$ 30 mJy, the predicted lensing fraction is $\sim92$ \%, increasing to nearly all sources with S$_{\rm{850 \mu m}}$ $>$ 40 mJy.

We rerun the same procedure on the 500 $\mu$m SPIRE fluxes, which shows that out of all 209 HerBS sources, we expect 158.1 $\pm$ 1.7 lensed sources, giving a total lensing fraction of 75.6 $\pm$ 0.8 per cent. This suggests that we are missing 29.7 $\pm$ 1.6 lensed sources with our SCUBA-2 observations.

Finally we note that our counts in the GAMA fields are systematically higher than those in the other H-ATLAS fields, a point also noticed by \cite{Negrello2016}. Using a similar method for the KS-test as described in Subsection \ref{sec:KS}, we calculate the probability for the GAMA and non-GAMA sources, and find a disagreement of 0.61 $\pm$ 0.47 standard deviations. This suggests the sources themselves do not differ significantly between the GAMA and the NGP+SGP fields.

% ++++++ Outlook ++++++
\section{Conclusions}
\label{sec:conclusions}
The HerBS catalogue consists of the brightest, high-redshift sources in the H-ATLAS survey, selected with S$_{\rm{500\mu m}} >$ 80 mJy and z$_{\rm{phot}} >$ 2. 
Initially, we selected 223 sources. 
SCUBA-2 observations of 203 of these sources allowed us to remove 14 blazars from the HerBS sample, leaving 20 HerBS sources unobserved.
152 out of the 189 confirmed high-redshift galaxies were detected at more than 3-$\sigma$, within 10 arc seconds of the SPIRE position. 
Currently, our \textit{Her}BS sample consists of 209 galaxies.

While recent studies like \cite{Scudder2016} suggest a significant effect of source confusion in \textit{Herschel} observations, none of our sources feature spatially-extended emission with $> 3\sigma$.
While some sources could be confused on a scale not probed by the SCUBA-2 observations, the lack of any signs at the detectable scales gives us little evidence of source confusion significantly affecting the purity of our sample. A reason for this could be due to our high lensing fraction, especially those caused by galaxy-galaxy lensing systems, whose influence is on a smaller angular scale than the less common galaxy-cluster lensing event.

We fitted a two-temperature blackbody as a template to the subset of 22 HerBS sources with spectroscopically determined redshifts, as well as to sub-samples where we divided our sources in redshift or luminosity. 
We find a cold- and hot-dust temperature of  $21.29_{-1.66}^{+1.35}$ K and $45.80_{-3.48}^{+2.88}$ K, a cold-to-hot dust mass ratio of $26.62_{-6.74}^{+5.61}$ and a $\beta$ of $1.83_{-0.28}^{+0.14}$.
%We find some significant differences between the templates that are derived by dividing the samples by redshift or by luminosity. However, as the sub-samples are sensitive to different ranges of the rest-frame wavelength, we cannot be sure the different templates are revealing any physical differences between the SEDs.
Overall, the fitted parameters are similar to previous work from \cite{Pearson2013}, and they agree broadly with the previous work from \cite{DunneEales2001,Clements2010}. We do not find evidence of any cold gas with temperatures below 20 K, as was found in \cite{PlanckThermal}. 

We find a high $\chi^2$ for the template, implying that the spectral energy distributions of the high-redshift population are diverse and cannot be represented by a single template. We showed that our improved template, which incorporates the SCUBA-2 flux densities, does not give a more accurate redshift estimates, which can also be explained by the diversity of the population.

Our sample has a similar redshift distribution as other samples selected at 500 $\mu$m, when we take the photometric redshift cut-off into account. Kolmogorov-Smirnov tests indicate that we probe a similar sample of galaxies as the SPT sample.

We calculated the number counts of the 850 $\mu$m observations of our sources, and compared them to a galaxy population model by \cite{Cai2013}. 
From this comparison we predict that 128.4 $\pm$ 2.1 out of the 152 SCUBA-2 detected, high-redshift galaxies are strongly lensed. A model based around the 500 $\mu$m flux suggests a total of 158.1 $\pm$ 1.7 of the 209 HerBS sources to be strongly lensed.
We report finding more lensed galaxies in the GAMA equatorial fields, when compared to the galaxy population model of \cite{Cai2013}, and the other fields (SGP + NGP).

\section*{Acknowledgements}

We would like to thank our JCMT operator, James Hoge, for his help at the full-service James Clerck Maxwell Telescope. TB, MS and SAE have received funding from the European Union Seventh Framework Programme ([FP7/2007-2013] [FP\&/2007-2011]) under grant agreement No. 607254. The \textit{Herschel}-ATLAS is a project with \textit{Herschel}, which is an ESA space observatory with science instruments provided by European-led Principal Investigator consortia and with important participation from NASA. The \textit{Herschel}-ATLAS website is \url{http://www.h-atlas.org}. RJI acknowledges support from ERC in the form of the Advanced Investigator Programme, 321302, COSMICISM. LD and SJM acknowledge support from ERC in the form of the Advanced Investigator Programme, 321302, COSMICISM, and the Consolidator Grant {\sc CosmicDust} (ERC-2014-CoG-647939, PI H,L,Gomez). MJM acknowledges the support of the National Science Centre, Poland through the POLONEZ grant 2015/19/P/ST9/04010. This project has received funding from the European Union's Horizon 2020 research and innovation programme under the Marie Sk{\l}odowska-Curie grant agreement No. 665778. MN acknowledges financial support from the European Union's Horizon 2020 research and innovation programme under the Marie Sk{\l}odowska-Curie grant agreement No 707601. H.D. acknowledges financial support from the Spanish Ministry of Economy and Competitiveness (MINECO) under the 2014 Ramón y Cajal program MINECO RYC-2014-15686. The authors thank the anonymous referee for his/her comments and suggestions.

%%%%%%%%%%%%%%%%%%%%%%%%%%%%%%%%%%%%%%%%%%%%%%%%%%



\bibliographystyle{mnras}

% if your bibtex file is called example.bib

%%%%%%%%%%%%%%%%%%%%%%%%%%%%%%%%%%%%%%%%%%%%%%%%%%

%%%%%%%%%%%%%%%%% APPENDICES %%%%%%%%%%%%%%%%%%%%%
\newpage
\appendix
\onecolumn
\captionsetup{width=0.9\linewidth}
\begin{landscape}
\section{HerBS Catalogue and Blazars}

\begin{longtable}{r r c c c c c c c c c c c c}

\caption{The HerBS sample - SPIRE and SCUBA-2 data. The HerBS number hyperlinks to the NED database at the position of the source. The RA and DEC are the SPIRE-positions, $\Delta$RA and $\Delta$DEC are the SPIRE positions minus the SCUBA-2 positions. Cursive SCUBA-2 observations are classed as non-detections, as discussed in Section \ref{sec:results}. The spectroscopic redshifts are discussed in Section \ref{sec:data}, z$_{\rm{phot,temp}}$ refers to the template derived in Section \ref{sec:data}, and z$_{\rm{phot,temp}}$ refers to the photometric redshift estimates in \citet{Ivison2015}. The bolometric luminosity is calculated using the fitted photometric template}S
\label{tab:AppendixA} \\
\multicolumn{1}{r}{No.} & \multicolumn{1}{r}{H-ATLAS ID} & \multicolumn{1}{c}{RA} & \multicolumn{1}{c}{DEC} &  \multicolumn{1}{c}{$\Delta$RA} & \multicolumn{1}{c}{$\Delta$DEC} & \multicolumn{1}{c}{S$_{250}$}  & \multicolumn{1}{c}{S$_{350}$} & \multicolumn{1}{c}{S$_{500}$} & \multicolumn{1}{c}{S$_{850}$} & \multicolumn{1}{c}{z$_{\rm{spec}}$} & \multicolumn{1}{c}{z$_{\rm{phot,temp}}$} &\multicolumn{1}{c}{z$_{\rm{phot,Ivi}}$} & \multicolumn{1}{c}{Lum.} \\ & & [hms] & [dms] & ["] & ["] & [mJy] & [mJy] & [mJy] & [mJy] & & & & log(L$_{\odot}$) \\ 
\hline \endhead
\hline 
\footnotemark[1]\href{http://ned.ipac.caltech.edu/cgi-bin/objsearch?search_type=Near+Position+Search&in_csys=Equatorial&in_equinox=J2000.0&lon=206.12285d&lat=30.509484d&radius=1.0&hconst=73&omegam=0.27&omegav=0.73&corr_z=1&z_constraint=Unconstrained&z_value1=&z_value2=&z_unit=z&ot_include=ANY&nmp_op=ANY&out_csys=Equatorial&out_equinox=J2000.0&obj_sort=Distance+to+search+center&of=pre_text&zv_breaker=30000.0&list_limit=5&img_stamp=YES}{1} & J134429.5+303034 & 206.1228 & 30.5095 & -1.4 & 2.04 & 461.9 $\pm$ 5.8 & 465.7 $\pm$ 6.5 & 343.3 $\pm$ 7.1 & 142.0 $\pm$ 8.1 & 2.30 & 2.21 & 2.33 & 13.94 \\
\href{http://ned.ipac.caltech.edu/cgi-bin/objsearch?search_type=Near+Position+Search&in_csys=Equatorial&in_equinox=J2000.0&lon=176.658163d&lat=-0.192264d&radius=1.0&hconst=73&omegam=0.27&omegav=0.73&corr_z=1&z_constraint=Unconstrained&z_value1=&z_value2=&z_unit=z&ot_include=ANY&nmp_op=ANY&out_csys=Equatorial&out_equinox=J2000.0&obj_sort=Distance+to+search+center&of=pre_text&zv_breaker=30000.0&list_limit=5&img_stamp=YES}{2} & J114637.9-001132 & 176.6582 & -0.1923 & -3.61 & 5.15 & 316.0 $\pm$ 6.6 & 357.9 $\pm$ 7.4 & 291.8 $\pm$ 7.7 & 148.1 $\pm$ 10.0 & 3.26 & 2.81 & 2.54 & 13.98 \\
\footnotemark[1]\href{http://ned.ipac.caltech.edu/cgi-bin/objsearch?search_type=Near+Position+Search&in_csys=Equatorial&in_equinox=J2000.0&lon=201.625499d&lat=33.735477d&radius=1.0&hconst=73&omegam=0.27&omegav=0.73&corr_z=1&z_constraint=Unconstrained&z_value1=&z_value2=&z_unit=z&ot_include=ANY&nmp_op=ANY&out_csys=Equatorial&out_equinox=J2000.0&obj_sort=Distance+to+search+center&of=pre_text&zv_breaker=30000.0&list_limit=5&img_stamp=YES}{3} & J132630.1+334408 & 201.6255 & 33.7355 & -2.11 & 4.02 & 190.5 $\pm$ 5.6 & 281.3 $\pm$ 5.9 & 278.6 $\pm$ 7.5 & 121.7 $\pm$ 8.7 & 2.95 & 3.77 & 3.21 & 14.08 \\
\end{longtable} 
\begin{longtable}{r r c c c c c c c c c c c c}
\caption{Blazars - SPIRE and SCUBA-2 data. The blazar index hyperlinks to the NED database at the position of the source. These sources have been removed from the HerBS sample in Section \ref{sec:measurements}. The RA and DEC are the SPIRE-positions, $\Delta$RA and $\Delta$DEC are the SPIRE positions minus the SCUBA-2 positions. The spectroscopic redshifts are discussed in Section \ref{sec:data}. The $\alpha$ value defines the steepness of the slope of the synchrotron radiation, and is calculated in Section \ref{sec:measurements}.}
\label{tab:AppendixBlazars} \\
\multicolumn{1}{r}{No.} & \multicolumn{1}{r}{H-ATLAS ID} & \multicolumn{1}{c}{RA} & \multicolumn{1}{c}{DEC} & \multicolumn{1}{c}{$\Delta$RA} & \multicolumn{1}{c}{$\Delta$DEC} & \multicolumn{1}{c}{S$_{250}$}  & \multicolumn{1}{c}{S$_{350}$} & \multicolumn{1}{c}{S$_{500}$} & \multicolumn{1}{c}{S$_{850}$} & \multicolumn{1}{c}{z$_{\rm{spec}}$} & \multicolumn{1}{c}{$\alpha$} \\ & & [hms] & [dms] & ["] & ["] & [mJy] & [mJy] & [mJy] & [mJy] & & \\ 
\hline \endhead 
\hline 
\href{http://ned.ipac.caltech.edu/cgi-bin/objsearch?search_type=Near+Position+Search&in_csys=Equatorial&in_equinox=J2000.0&lon=197.619669d&lat=32.345513d&radius=1.0&hconst=73&omegam=0.27&omegav=0.73&corr_z=1&z_constraint=Unconstrained&z_value1=&z_value2=&z_unit=z&ot_include=ANY&nmp_op=ANY&out_csys=Equatorial&out_equinox=J2000.0&obj_sort=Distance+to+search+center&of=pre_text&zv_breaker=30000.0&list_limit=5&img_stamp=YES}{1} & J131028.7+322044 & 197.6197 & 32.3455 & -2.06 & 2.05 & 259.1 $\pm$ 4.4 & 363.1 $\pm$ 5.1 & 452.2 $\pm$ 6.1 & 820.0 $\pm$ 6.0 & 1.0 & 0.93 $\pm$ 0.01 \\
\href{http://ned.ipac.caltech.edu/cgi-bin/objsearch?search_type=Near+Position+Search&in_csys=Equatorial&in_equinox=J2000.0&lon=137.292356d&lat=1.359685d&radius=1.0&hconst=73&omegam=0.27&omegav=0.73&corr_z=1&z_constraint=Unconstrained&z_value1=&z_value2=&z_unit=z&ot_include=ANY&nmp_op=ANY&out_csys=Equatorial&out_equinox=J2000.0&obj_sort=Distance+to+search+center&of=pre_text&zv_breaker=30000.0&list_limit=5&img_stamp=YES}{2} & J114637.9-001132 & 137.2924 & 1.3597 & -0.52 & -0.03 & 256.5 $\pm$ 3.8 & 327.0 $\pm$ 4.5 & 375.3 $\pm$ 6.0 & 390.7 $\pm$ 7.7 & 1.02 & 0.32 $\pm$ 0.01 \\
\href{http://ned.ipac.caltech.edu/cgi-bin/objsearch?search_type=Near+Position+Search&in_csys=Equatorial&in_equinox=J2000.0&lon=26.26398d&lat=-27.559145d&radius=1.0&hconst=73&omegam=0.27&omegav=0.73&corr_z=1&z_constraint=Unconstrained&z_value1=&z_value2=&z_unit=z&ot_include=ANY&nmp_op=ANY&out_csys=Equatorial&out_equinox=J2000.0&obj_sort=Distance+to+search+center&of=pre_text&zv_breaker=30000.0&list_limit=5&img_stamp=YES}{3} & J014503.4-273333 & 26.264 & -27.5591 & -0.67 & 0.98 & 131.5 $\pm$ 5.7 & 179.1 $\pm$ 6.3 & 233.5 $\pm$ 7.2 & 365.8 $\pm$ 6.7 & 1.16 & 0.83 $\pm$ 0.02 \\
\end{longtable}

\end{landscape}
\section{Cutouts of the entire HerBS sample}
\label{sec:AppC}
\begin{center}
\begin{figure} 
%This needs to be expanded in the online version, to accommodate all 27 files
	\includegraphics[width=0.9\columnwidth]{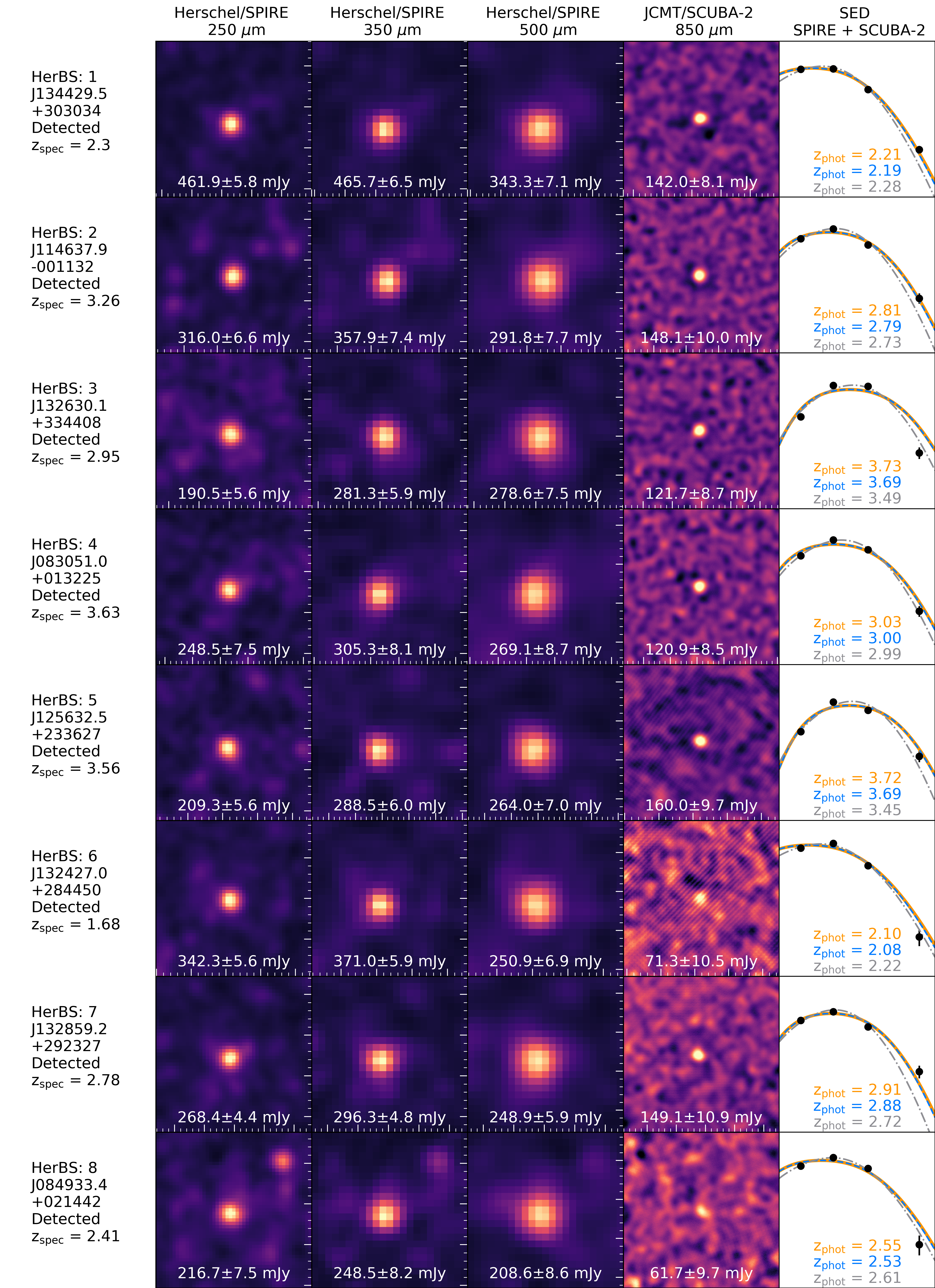}
\end{figure}
\end{center}

%%%%%%%%%%%%%%%%%%%%%%%%%%%%%%%%%%%%%%%%%%%%%%%%%%

% Don't change these lines
\bsp	% typesetting comment
\label{lastpage}
\end{document}